\documentclass[journal,draftcls,onecolumn,12pt,twoside]{IEEEtranTCOM}

\normalsize
\usepackage{color}
\usepackage{amsmath,amssymb,amsfonts,amsthm,mathrsfs,bm}
\usepackage{graphicx}
\usepackage{balance}
\usepackage{setspace}
\usepackage{cite}
\usepackage{graphics}
\usepackage{graphicx}
\usepackage{subfigure}
\usepackage{caption}
\usepackage{algorithm}               
\usepackage{algorithmic}             
\usepackage{multirow} 
\usepackage{sidecap}
\usepackage{amsmath}

\renewcommand{\algorithmicrequire}{\textbf{Input:}} 
\renewcommand{\algorithmicensure}{\textbf{Output:}} 

\makeatletter
\renewcommand\normalsize{%
	\abovedisplayskip 2.5\p@ \@plus3\p@ \@minus4\p@
	\abovedisplayshortskip \z@ \@plus4\p@
	\belowdisplayshortskip 2.5\p@ \@plus2\p@ \@minus0\p@
	\belowdisplayskip \abovedisplayskip
	\let\@listi\@listI}
\makeatother
\newtheorem{theorem}{Theorem}
\newtheorem{lemma}{Lemma}

\newtheorem{definition}{Definition}

\theoremstyle{remark}
\theoremstyle{corollary}
\theoremstyle{definition}
\allowdisplaybreaks

\begin{document}

\title{Joint Queue-Aware and Channel-Aware Delay Optimal Scheduling of Arbitrarily Bursty Traffic over Multi-State Time-Varying Channels}
%
%
%

\author{Meng~Wang, Juan~Liu, Wei~Chen,~\IEEEmembership{Senior Member, IEEE,} Anthony~Ephremides,~\IEEEmembership{Life~Fellow,~IEEE}

}


\maketitle
\begin{spacing}{1.5}
\begin{abstract}
This paper is motivated by the observation that the average queueing delay can be decreased by sacrificing power efficiency in wireless communications. In this sense, we naturally wonder what is the minimum queueing delay when the available power is limited and how to achieve the minimum queueing delay. To answer these two questions in the scenario where randomly arriving packets are transmitted over multi-state wireless fading channel, a probabilistic cross-layer scheduling policy is proposed in this paper, and characterized by a constrained Markov Decision Process (MDP). Using the steady-state probability of the underlying Markov chain, we are able to derive the mathematical expressions of the concerned metrics, namely, the average queueing delay and the average power consumption. To describe the delay-power tradeoff, we formulate a non-linear programming problem, which, however, is very challenging to solve. By analyzing its structure, this optimization problem can be converted into an equivalent Linear Programming (LP) problem via variable substitution, which allows us to derive the optimal delay-power tradeoff as well as the optimal scheduling policy. The optimal scheduling policy turns out to be dual-threshold-based, which means transmission decisions should be made based on the optimal thresholds imposed on the queue length and the channel state.
\end{abstract}


%
\IEEEpeerreviewmaketitle


\section{Introduction}
%
%
%
%
\IEEEPARstart{W}{ith} the explosive proliferation of mobile devices, future wireless networks should provide an increasing number of multimedia applications  with more stringent Qualities of Service (QoS). Among various QoS metrics, latency and energy efficiency are two key metrics of interest\cite{6824752,6157574}. Low latency is highly expected when providing delay-sensitive or time-critical applications such as in Tactile Internet \cite{fettweis2014tactile} and is an important metric in URLLC (Ultra-Reliable Low Latency Communications) which is a new feature brought by 5G \cite{Popovski2017Ultra}. Meanwhile, high energy efficiency is urgently required especially for mobile devices powered by rechargeable batteries of finite capacity. Thus, it is of great importance to study the delay optimal and energy/power efficient transmission strategy for users in wireless communications\cite{1365285,6802871}. Intuitively, to reduce the latency, the transmitter would conduct transmission more frequently or increase the transmission data rate, which inevitably costs more transmission power. Therefore, there exists a fundamental tradeoff between the average queueing delay and the average power/energy consumption. 

In general, it is very challenging to derive the delay-power tradeoff in wireless communication systems, considering the randomness of data packet arrivals, and the time-varying characteristics of wireless channels. These randomness occur in different layers of the transmitter, which increases the difficulty of characterizing the delay-power tradeoff\cite{720543}. To deal with this issue, the cross-layer design framework, first presented in \cite{collins1999transmission}, was proposed to capture the uncertainties occurring at different layers in the last decades\cite{berry2002communication,berry2004cross,6482230,6152121,6477062,1425747}.

Within the cross-layer architecture, many works have focused on revealing the delay-power tradeoff, which can be classified into two major categories. One line of the works attempt to find the analytical delay-power tradeoffs by considering some ideal or simplified  assumptions on the system model \cite{uysal2002energy,zafer2009calculus,5510780,chen2007optimal,6213038}. In \cite{uysal2002energy}, the authors proposed a scheduling policy named Lazy scheduling which assigns transmission chances based on the backlog in the queue under the assumption that the arrival times of the packets are known in advance. In \cite{zafer2009calculus}, the authors minimized the transmission power with QoS constraits by assuming that the data arrival is known ahead of schedule and the channel is static or slow fading. The power constrained delay minimization problem was studied for a cognitive multi-access channel and a two-state block fading channel in \cite{5510780} and \cite{chen2007optimal}, respectively. 
This line of works mainly provide theoretical value more than engineering value, since the assumptions are too ideal to be practical. However, they are able to provide deeper insights to guide for engineering applications such as protocol design. 

\begin{figure}[t]
	\centering
	\includegraphics[width=0.5\columnwidth]{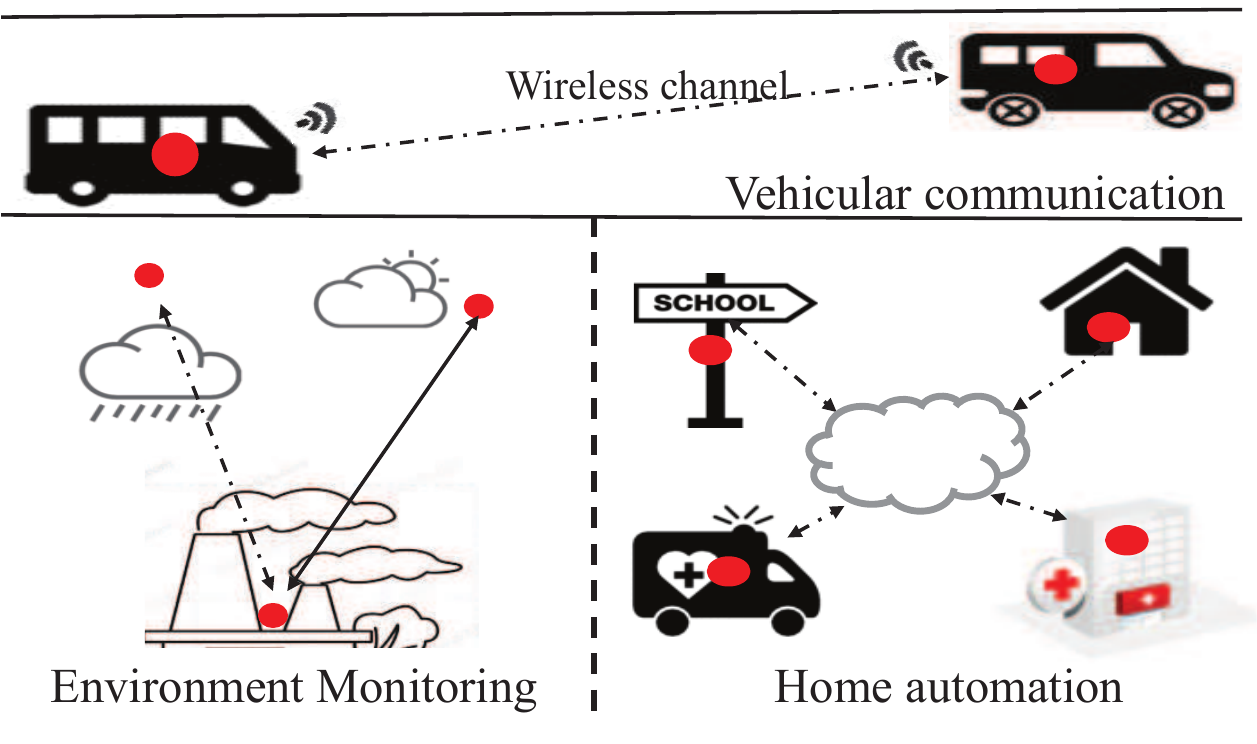}
	\caption{Potential application scenarios in sensor network}
	\label{model-1}
	\vspace{-1cm}
\end{figure}

The works in the other category consider more complex and practical system models\cite{berry2002communication,berry2004cross,6482230,ata2005dynamic,1262621,1315903}. In \cite{berry2002communication}, Berry and Gallager proposed  adapting the users' transmission rate and power by regulating the average power and average buffer delay over a wireless fading channel. They also focused on studying the cross-layer resource allocation in wireless fading channels for \cite{berry2004cross} and deriving the optimal power-delay tradeoff for a single user in the regime of asymptotically small delays in \cite{6482230}. Ata investigated the power minimization problem subject to the packet drop rate in \cite{ata2005dynamic}, assuming the fixed channel state, Possion packet arrival and exponentially distributed packet size. In \cite{1262621,1315903}, the authors studied the delay-bounded packet scheduling problem with bursty traffic arrival over wireless channels. This line of works studied the delay-power curve and analyzed its property under some circumstances. While it is difficult to derive theoretical solutions in general cases. This line of works mainly focus on studying resource allocation solutions and designing efficient algorithms for practical usage, which is of great importance in designing delay/power-efficient wireless transmission strategies.


More recently, 
a simple probabilistic scheduling policy was proposed to achieve the minimum queueing delay under power constraint in our previous work \cite{chen2007optimal}, where Bernoulli packet arrivals and a two-state fading channel model were considered. Further, arbitrarily random packet arrival patterns were considered to capture the impact of bursty network traffic in \cite{7417380,19920626} and adaptive transmission is considered in \cite{7893801}. In these works, we proved that the optimal delay-power tradeoff can be achieved by applying the optimal scheduling polices which determine packet transmissions based on the threshold imposed on the queue length. The structured policy is appealing for the scheduler thanks to its ease of  deployment
. Hence, it inspires us to further dig into this topic. We naturally wonder if the optimal solution still has a special structure in more general scenarios and what kind of structure it may have.

In this paper, we study the delay-power tradeoff in wireless packet transmissions in a more realistic but complex communication system, where data packets are generated from an arbitrarily bursty traffic and a multi-state wireless fading channel is considered. Some potential application scenarios are shown in \figurename\ref{model-1}. The major challenges of this work lie in two aspects: 1) how to perform probabilistic scheduling jointly based on the randomness of the data packet arrival, the occupancy of the transmission data queue, and the time-varying characteristics of the wireless channel, and 2) how to reveal the structure of the optimal policy.

At the first task, the major challenge confronted is to build a proper cross-layer framework which includes all the system dynamics. Incorporating all these effects, our proposed scheduling policy performs joint scheduling based on the time-varying environment. Hence, it is very challenging to formulate the optimal cross-layer scheduling problem while facilitating theoretical analysis of its optimal solution. To deal with this difficulty, we propose a stochastic scheduling policy being aware of packet arrival, buffer and channel states. Then, we formulate a non-linear optimization problem to find the optimal probabilistic scheduling parameters. The challenge behind the second task is how to solve the optimal scheduling problem and derive the closed-form solution. This lies in the fact that the dimensionality of solving the optimal scheduling problem increases significantly due to the enlarged number of scheduling parameters that increases linearly with the number of channel and packet arrival states. By solving the obtained non-linear problem, we can surely obtain the optimal delay-power tradeoff. However, it is not trivial to search for the optimal solution to the non-linear optimization problem, let alone derive the optimal scheduling solution theoretically. To deal with this challenge, we first find a method to convert it to an LP problem, through which we can further analyze the structure of the optimal solution and reveal that the optimal scheduling policy has a dual-threshold-based structure step by step. By dual-threshold-based, we mean that packets should be transmitted based on the thresholds imposed on not only the queue state but also the on channel state. 

The remainder of this paper is organized as follows. The system setting is introduced in Section \ref{sec2}. In Section \ref{sec3}, we propose the probabilistic scheduling policy to schedule packet transmissions based on the buffer and the channel states simultaneously. In Section \ref{sec4}, 
we formulate a non-linear power constrained delay minimization problem and then convert it to an equivalent LP problem. In Section \ref{sec5}, we reveal that the optimal scheduling policy is dual-threshold-based with a rigorous mathematic proof and propose an algorithm to find simplified suboptimal policy. Simulation results are demonstrated in Section \ref{sec6} to validate the dual-threshold-based policy and concluding remarks are presented in Section \ref{sec7}. Some notations frequently used are explained as follows. Given a positive integer $K$, the notation  $\mathbb{K}$ denotes an integer set $\{0,1,2,\cdots,K\}$ while $\mathbb{K}^+$ denotes integer set $\mathbb{K}/\{0\}$. Sets $\mathbb{W}$ and $\mathbb{W}^+$, $\mathbb{M}$ and $\mathbb{M}^+$ are defined in the same way.\footnote{Part of this work was published in \cite{7848871}, where main results were presented while most important derivations for some conclusions towards the dual-threshold-based structure were omitted due to the limited space.}

\section{System Model} \label{sec2}

\begin{figure}[t]
	\centering
	\includegraphics[width=0.7\columnwidth]{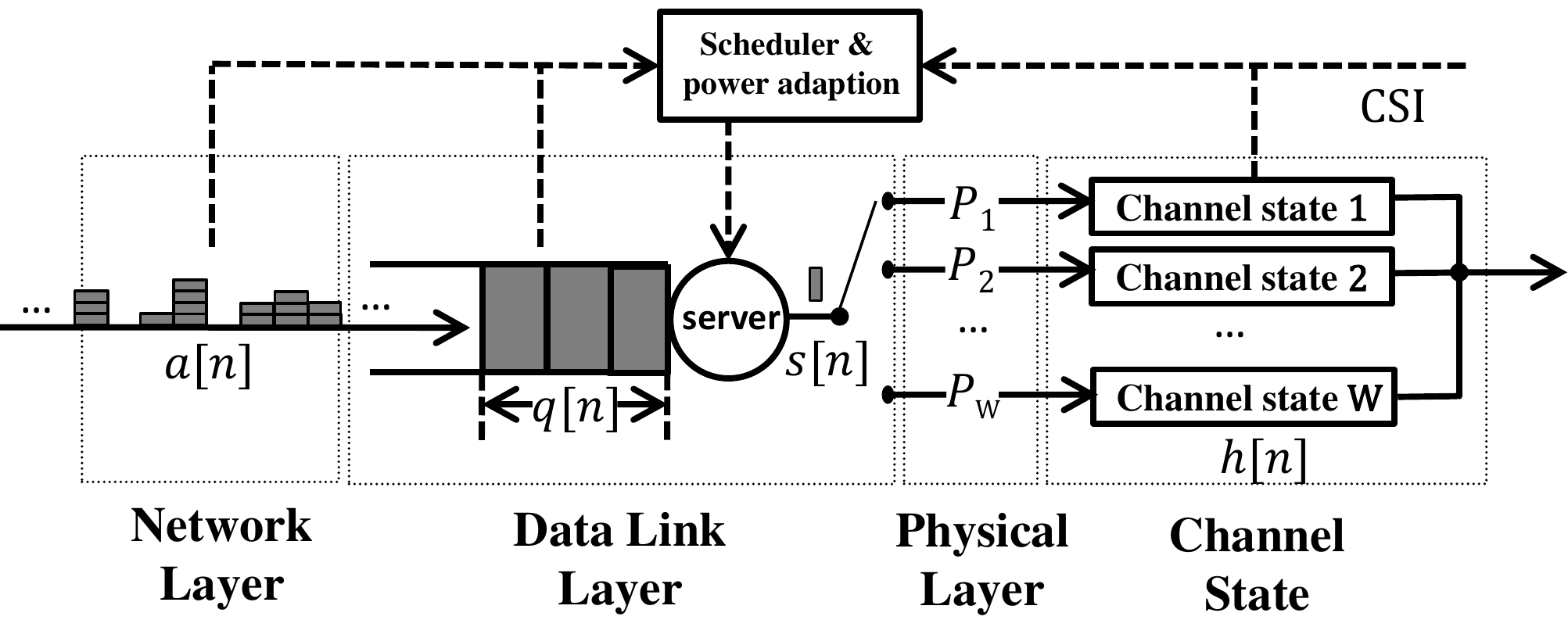}
	\caption{System model}
	\label{model_2}
	\vspace{-1cm}
\end{figure}

We consider a wireless communication system where the source node transmits to the destination over a time-varying wireless link. As shown in \figurename\ref{model_2}, packets of bursty traffic generated by higher-layer applications arrive at the network layer randomly, and are stored at the buffer in the data link layer. In the physical layer, the transmitter determines when to transmit the queued packets over a multi-state wireless channel, with the aid of efficient scheduling policies.

Let $a[n]$ denote the number of packets randomly arriving in the $n$th slot. To capture the burstiness and variability of real-time applications, we assume an arbitrarily packet arrival pattern, i.e., the number of newly arriving packets could follow any distribution. Suppose that $a[n]$ is independent and identically distributed ($i.i.d.$). Thus, the mass probability function of $a[n]$ can be characterized by
\begin{align}
\text{Pr}\{a[n]=m\}=\theta_m, m = 0,1,2,\cdots
\end{align}
where $\theta_m \in [0,1]$. Considering traffic shaping and admission control adopted in the system, the number of packets newly arriving in each time slot must be upper-bounded by a large integer $M$. In other words, there exists a positive  integer $M$ such that $\theta_m = 0$, for all $m>M$, and $\sum_{m=0}^{M} \theta_m=1$. The average packet arrival rate $\bar{a}$ is obtained as
\begin{align}\label{eq2}
\bar{a}=\lim\limits_{N\rightarrow\infty}sup \ \frac{1}{N}\sum\limits_{n=0}^N a[n] =\sum\limits_{m=0}^M m\cdot \theta_m.
\end{align}

At the source node, a buffer is employed to store the backlogged packets which cannot be sent immediately. 
The queue state, denoted by $q[n]$, is characterized by the number of packets in the buffer at the end of $n$th slot and updated as
\begin{equation}\label{eq3}
\begin{split}
q[n]&=max\big\{ min\{q[n\!-\!1]+a[n], K\}-s[n] , 0 \big\},\\
\end{split}
\end{equation}
where $s[n]$ is the transmitted packets in the $n$th time slot and $K$ is the capacity of the buffer\footnote{Packet overflow will occur if $K$ is quite small. In this work, we assume that $K$ is a sufficiently large constant such that no packet overflow will occur. In Section V, we give the conclusion that if $K$ is greater than a threshold, the queueing length will never reach the capacity according to our proposed scheduling scheme. Thus, the max operation in \eqref{eq3} can be omitted.}.

We adopt a $W$-state block fading channel model, where $W$ is a positive integer. Let $h[n]$ denote the channel state in the $n$th time slot. By  block fading, we mean that the channel state $h[n]$ stays invariant during each time slot and follows an $i.i.d.$ fading process across the time slots. Here, the discrete $W$ channel states indicate different wireless channel qualities. Let  $d_1 = 0 < d_2 <\cdots < d_W < d_{W+1} = \infty$ be the channel power gain levels. If the channel gain in the $n$th time slot ranges in interval $[d_w,d_{w+1})$, we say that the wireless channel is at state $w$. 
Since the channel quality becomes better with the increase of the index, $w = 1$ and $w = W$ represent the worst and the best channel condition, respectively. 
The mass probability function of $h[n]$ is described as
\begin{align}
\text{Pr}\big\{h[n] = w \big \}=\eta_w,
\end{align}
where $\eta_w \in [0,1]$ and $ w\in \mathbb{W}^+$.


Suppose that there exists a feedback channel through which the Channel State Information (CSI) is sent back from the receiver to the transmitter. Intuitively, the transmission power shall be adapted to the channel state to meet the requirement of successful packet delivery. Let $P_w$ ($ w\in \mathbb{W}^+$) denote the power needed to transmit one packet successfully in the channel sate $w$. Since more power is required to combat wireless channel fading when the channel condition is worse, it is reasonable to assume $P_1>P_2>\cdots>P_w>\cdots>P_W$.

In our model, we consider a fixed-rate transmission scheme which has been widely adopted in practice \cite{qiao2009impact}. Without loss of generality, we assume the transmission rate is one packet per slot. Hence, at most one data packet can be delivered in each slot, namely, $s[n]$ $\in$ $\{0,1\}$.

In the cross-layer design framework shown in \figurename\ref{model_2}, the scheduler will schedule packets transmissions in each slot $n$ based on the packet arrival state $a[n]$, the queueing state $q[n-1]$, and the channel state $h[n]$ subjected to a power constraint, as will be discussed in details in the next section where the scheduling problem is treated as a power constrained Markov Decision Process (MDP), and discussed in Section \ref{sec4}.

\section{Probabilistic Scheduling Policy}\label{sec3}

In this section, we introduce a probabilistic scheduling policy based on which the transmitter decides whether or not to deliver one data packet to its receiver in each slot.

\subsection{Probabilistic Scheduling}\label{scheduling policy}

To improve the power efficiency, the transmitter should exploit a better channel state to deliver the packets to spend much less power. Thus, the source is more willing to keep silent till the channel state gets better. However, this may induce undesirable large latency waiting for good channel states, which is intolerable for serving delay-sensitive or time-critical traffics. To overcome this issue, some backlogged packets should be transmitted immediately at the cost of consuming higher power, even when the channel state may not be so good. Hence, the proposed scheduler must achieve a balance between the average delay and the power consumption.

In this work, a probabilistic cross-layer scheduling policy is proposed to schedule packet transmissions in each time slot. At the beginning of the $n$th time slot, the scheduler collects the current system state including the queueing state $q[n-1]=k$, the packet arrival state $a[n]=m$, and the channel state $h[n]=w$. Given $q[n-1]=k$, $a[n]=m$, and $h[n]=w$, it decides to transmit one packet with probability $f_{k+m,w}$ or keep silent with probability $1-f_{k+m,w}$
. By $f_{k+m,w}$, we mean that the scheduler can schedule packet transmissions based on the updated queue state $q[n-1]+a[n]=k+m$ after one packet arrival. The reason lies in the fact that one of the packets newly arriving at this slot can be delivered immediately. Hence, it is not necessary to distinguish between the backlogged packets and the newly arriving packets.  Clearly, the transmission probability $f_{k+m,w}$ lies in the interval  $[0,1]$. 

According to the above probabilistic scheduling policy, the number of transmitted packets $s[n]$ for the current slot is a random variable, the probability mass function of which is given by 
\begin{align}\label{eq5}
s[n]=\left\{
\begin{array}{ll}
1  \quad w.p. \quad f_{k+m,w},\\
0  \quad w.p. \quad 1-f_{k+m,w},
\end{array}
\right.
\end{align}
where $k \in \mathbb{K}, m \in \mathbb{M}, w \in \mathbb{W}$ and the abbreviation $'w.p.'$ is short for $'with\ the \  probability \ of'$\footnote{In Eq. \eqref{eq5}, when $a[n]+q[n-1]=0$, there is no packet waiting to be transmitted,  and when $a[n]+q[n-1]>K$, packet loss will happen. Thus, $f_{0,w}$ ($w \in \mathbb{W}$) and $f_{k+m,w}$ ($k+m>K,\ w \in \mathbb{W}$) are set as zero for notational consistence.}.


We aim to find the optimal policy with a set of optimal transmission probabilities $\{f_{k+m,w}^*\}$  that can minimize the average queueing delay under an average transmission power constraint.

\begin{figure*}[t]
	\centering
	\includegraphics[width=0.9\columnwidth]{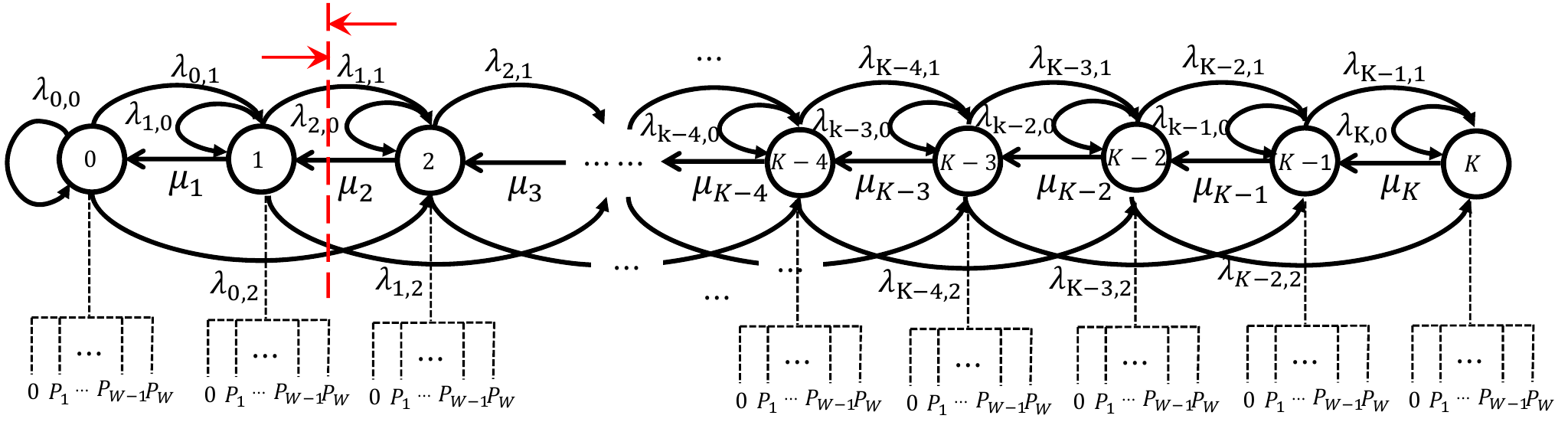}
	\caption{Illustrative of the MDP model with $M=2$: for each queue length, a transmission power $P_w$ is consumed when transmitting one packet over the channel state $w$.}
	\label{mdp}
\end{figure*}

\subsection{Markov Decision Process} \label{steady-proc}

Based on the scheduling policy in section \ref{scheduling policy}, the scheduler makes decision of transmitting $s[n]$ packet(s) in every slot. The transmission decision affects the number of the packets queueing in the buffer as well as the transmission power. In this sense, we model the scheduling problem as a constrained MDP with the queue length $q[n]$ being the system state. The decision, either waiting or transmitting ($s[n]\in\{0,1\}$), is treated as one candidate action taken at the current state. Executing each action certainly causes some system costs, namely, the delay cost associated with the queue length and the power cost associated with the packet transmission. Let $\tau_{k,l}$ denote the one-step state transition probability from state $q[n-1]=k$ to state $q[n]=l$, i.e., 
\begin{align}
\tau_{k,l} =\text{Pr}\big\{q[n]=l~|~q[n-1]=k\big\}.
\end{align}
The transition probabilities of the underlying Markov chain are presented in \emph{Lemma} \ref{theorem1}.

\begin{lemma}\label{theorem1}
	The forward and backward state transition probabilities denoted by $\lambda_{k,m}=\tau_{k,k+m}$ and $\mu_k=\tau_{k,k-1}$ are obtained as
	\vspace{-0.0cm}
	\begin{align}
	\lambda_{k,m}\!& = \theta_m\sum\nolimits_{w=1}^W \eta_w(1-f_{k+m,w}) + \theta_{m+1}\sum\nolimits_{w=1}^W \eta_w f_{k+m+1,w} ,\label{eq601}
	\end{align}
	\begin{align}
	\mu_k& = \theta_0\sum\nolimits_{w=1}^W \eta_w f_{k,w},\label{eq602}
	\end{align}
	\vspace{-0.0cm}
	where $k \in \mathbb{K}$ and $m \in \mathbb{M}^+$. The state transition probability $\lambda_{k,0}$ is the probability that the queue length remains the same, given by
	\vspace{-0.0cm}
	\begin{align}\label{eqk0}
	\lambda_{k,0}=\tau_{k,k}=\left\{
	\begin{array}{ll}
	1 \!-\! \sum\nolimits_{m=1}^M\lambda_{k,m}, &  k=0,\\
	1\! -\! \sum\nolimits_{m=1}^{M}\lambda_{k,m} \!-\! \mu_k,  &  k \in \mathbb{K}^+.
	\end{array}
	\right.
	\end{align}
\end{lemma}
\begin{IEEEproof}
From Eq. \eqref{eq3}, the transition from state $q[n-1]=k$ to $q[n]=k+m$ takes place with probability $\lambda_{k,m}$ when $a[n]-s[n]=m$.\footnote{Due to the assumption in Footnote $1$, the operators "max" and "min" in Eq. \eqref{eq3} can be omitted here.} This happens in two cases. In Case I, when there are $a[n]=m$ packets arrive at the queue over channel state $w$, no packet is delivered ($s[n]=0$) with probability $1-f_{k+m,w}$. In this situation, the transition probability is $\theta_m(1-f_{k+m,w})$.In case II, when there is $a[n]=m+1$ packets arrive at the queue over channel state $w$, one packet is transmitted ($s[n]=1$) with probability $f_{k+m+1,w}$. Accordingly, the transition probability is $\theta_{m+1}f_{k+m+1,w}$. Combining these two cases,  $\lambda_{k,m}$ can be calculated using the law of total probability, and shown in Eq. \eqref{eq601}. Similarly, the probability $\mu_k$ can be derived, as given by Eq. \eqref{eq602}. The probability $\lambda_{k,0}$ given by Eq. \eqref{eqk0} is obtained using probability normalization.
\end{IEEEproof}
Notice that, $\tau_{k,l}=0$ holds for $|l\!-\!k|>M$, since the queue length increases from $k$ up to $l=k+M$ after one packet arrival. In \figurename \ref{mdp}, we show an example of the MDP model with $M=2$. In each time slot, $q[n]$ increases by no more than $M$ due to one new data arrival, while decreases by one since at most one packet can be delivered. Let matrix $\mathbf{\Lambda}$ denote the $(K\!+\!1)$-by-$(K\!+\!1)$ transition probability matrix of the underlying Markov chain. The $(j\!+\!1,i\!+\!1)$-th element of $\mathbf{\Lambda}$ is transition probability $\tau_{i,j}$. The transition probability matrix $\mathbf{\Lambda}$ is a banded matrix, since the number of the newly arrival packets and departing packets are limited in one slot.

Let $\pi_{k}$ denote the steady-state probability of the queue length being equal to $k$. The stationary distribution of the system state is denoted by the vector $\bm{\pi} = [\pi_0,\pi_1,\cdots, \pi_K]^T$, where the superscript $T$ denotes matrix transpose. Vectors $\mathbf{1}$ and $\mathbf{0}$ are used to denote the $(K+1)$-dimensional column vectors whose entries are zero and one, respectively. According to the property of the steady-state probability, we have $\mathbf{\Lambda}\bm{\pi}=\bm{\pi}$ and $\mathbf{1}^T \bm{\pi} =1$. Hence, the stationary distribution $\bm{\pi}$ is the solution to the following linear equations
\begin{align}\label{steady-pi}
\left[
\begin{array}{c}
\mathbf{Q}_K\\
\mathbf{1}^T
\end{array}
\right]\bm{\pi} =
\left[
\begin{array}{c}
\bm{0}\\
1
\end{array}
\right],
\end{align}
where $\mathbf{Q}_K$ is a matrix consisting of the first $K$ rows of the generator matrix $\mathbf{Q} = \mathbf{\Lambda} - \mathbf{I}$. From Eq. \eqref{steady-pi} and \emph{Lemma} \ref{theorem1}, we can see that the steady-state probability $\bm{\pi}$ is determined by the scheduling policy with the parameters $\{f_{k+m,w}\}$.

\section{Delay and Power Tradeoff}\label{sec4}

In this section, we first analyze the two key performance metrics: the average queueing delay and the average power consumption. Then, we formulate optimization problems to describe the delay minimum power constrained scheduling problem, based on the stationary probability of the built Markov Decision Process.

\subsection{Delay and Power Metrics}\label{Delay and Power Metrics}

In accordance with every transmission action $s[n]$, the scheduler spends some system costs due to queue occupation and packet transmission. Given action $s[n]$, the queueing cost for buffer occupation is denoted by $C_q[n]$ and the power cost for packet transmission is denoted by $C_p[n]$, respectively, expressed as
\begin{align}
C_q[n]=(q[n-1]+a[n]-s[n])^+ ~\text{and} ~C_p[n]=P_w s[n] \label{11102138}
\end{align}
As time goes by, the time-average costs can be built up as 
\begin{align}
Q_{\Omega} = \lim_{N\rightarrow\infty}\frac{1}{N}\sum\limits_{n=1}^{N}C_q[n]~~ \text{and}~~
P_{\Omega} = \lim_{N\rightarrow\infty}\frac{1}{N}\sum\limits_{n=1}^{N}C_p[n], 
\end{align}
respectively. Considering  the minus and connotative plus operators before $s[n]$ in Eq. \eqref{11102138}, an action $s[n]$ exerts opposite influences on the buffer occupation and power consumption, which naturally leads to a tradeoff between the average  delay (the Little's Law) and the average power.

The above analyses explain the average delay and the average power from the cost perspective of the scheduling policy. It's much easier to understand the tradeoff from the expressions of the two metrics given in Eq. \eqref{11102138}. To mathematically derive the two metrics, we refer to the MDP model built in Section \ref{steady-proc}. Once the stationary distribution $\bm{\pi}$ is obtained, the average queueing delay and power consumption can be derived and shown in the following theorem. 
\begin{theorem} \label{DP}
	Given a probabilistic scheduling policy $\{f_{k+m,w}\}$, the average queueing delay $D$ and power consumption $P$ can be expressed as
	\begin{align}
	&D = \frac{1}{\bar{a}}\sum\limits_{k=0}^{K} k\pi_k, \label{eq10}\\
	&P = \sum\limits_{k=0}^K  \pi_k \sum\limits_{w=1}^W  \eta_w P_w \sum\limits_{m=0}^M   \theta_m f_{k\!+\!m,w}. \label{eq12}
	\end{align}
\end{theorem}
\begin{IEEEproof}
	Given the stationary probability distribution of the Markov chain, the average queue length can be expressed as $Q = \mathbb{E}\{q[n]\}= \sum_{k=0}^{K} k\pi_k$.
	Then according to the Little's Law \cite{kleinrock1975queueing}, the average queueing delay $D$ can be derived as $Q/\bar{a}$ and shown in Eq. \eqref{eq10}. 
	
	With $C_p[n] = P_ws[n]$, we have $C_p[n]=P_w$ and $C_p[n]=P_0=0$, respectively, when one packet is transmitted over the channel state $w$, i.e., $s[n]=1$, and no transmission takes place, i.e., $s[n]=0$. Let $\psi_{k,w}$ denote the conditional probability of $C_p[n]=P_w$ $\big(w \in \mathbb{W} = \{0 \cup \mathbb{W}^+\}\big)$ given the queue state $q[n\!-\!1]\!=\!k$ and channel state $h[n]\!=\! w$. It can be expressed as
	\begin{align}\label{eq11}
	\psi_{k,w} =& \text{Pr} \{ C_p[n]\!=\!P_w \big| q[n\!-\!1]\!=\!k,h[n]\!=\! w \} 
	=  \left\{
	\begin{array}{ll}
	\sum\limits_{m=0}^{M} \theta_m f_{k+m,w}, & w \in \mathbb{W}^+, \\
	1-\sum\limits_{w=1}^{W} \psi_{k,w}, & w = 0.
	\end{array}
	\right.
	\end{align}
	By the law of total probability, the average power can be derived as
	\begin{align}
	\small
	{P}= & \sum\limits_{k\!=\!0}^K \sum\limits_{w\!=\!1}^W  \text{Pr}\{q[n\!-\!1]=k\}\text{Pr}\{h[n]=w\} \text{Pr} \{ C_p[n]\!=\!P_w \big| q[n\!-\!1]\!=\!k,h[n]\!=w \} \times P_w \nonumber\\
	=  &\sum\limits_{k=0}^K \sum\limits_{w=1}^W  \pi_k \eta_w \psi_{k,w}P_w  \\
	= & \sum\limits_{k=0}^K  \pi_k \sum\limits_{w=1}^W  \eta_w P_w \sum\limits_{m=0}^M   \theta_m f_{k\!+\!m,w}. \nonumber
	\end{align}
\end{IEEEproof}

We notice that, the steady state probability $\bm{\pi}$ is an  implicit function of the transmission probabilities, since it is uniquely determined by the transmission probabilities $\{f_{k+m,w}\}$ based on the analyses in Section \ref{steady-proc}. Thus, from \emph{Theorem} \ref{DP}, the average queueing delay and the average power consumption are both functions of transmission probabilities.

\subsection{Delay-Power Tradeoff}

To find the optimal scheduling policy with a set of transmission probabilities $\{f_{k+m,w}^* | , k \in \mathbb{K}, m \in \mathbb{M}, w \in \mathbb{W}\}$, we formulate an optimization problem to minimize the average queueing delay $D$ under the power constraint $P_{aver}$ as follows:
\begin{equation}\label{tradeoff1}
\begin{split}
&\mathop{\text{min}}\nolimits_{\{f_{k+m,w} \}} \  D=\frac{1}{\bar{a}}\sum\nolimits_{k=0}^{K} k\pi_k \\
&\text{s.t.}
\left\{
\begin{array}{ll}
P  \leqslant P_{aver}   & (a)\\
f_{k+m,w} \in [0,1],  & (b)\\
\mathbf{Q}\bm{\pi}=\bm{0} &(c)\\
\mathbf{1}^T \bm{\pi} =1  & (d)\\
\bm{0} \preceq \bm{\pi} \preceq \bm{1} &(e)
\end{array}
\right.
\end{split}
\end{equation}
where $k \in \mathbb{K},~m \in \mathbb{M}, ~w \in \mathbb{W}$, and symbol '$\preceq$' represents the component-wise inequality between vectors. In problem \eqref{tradeoff1}, the objective is to minimize the average queueing delay. Constraint (\ref{tradeoff1}.a) denotes the maximum power constraint. Constraint (\ref{tradeoff1}.b) indicates the range of the optimization variables $\{f_{k+m,w}\}$. Constraints (\ref{tradeoff1}.c-\ref{tradeoff1}.e) are derived from the properties of the Markov chain. Constraint (\ref{tradeoff1}.e) specifies the range of the steady-state probabilities. Since problem \eqref{tradeoff1} is a non-linear programming problem, it is rather difficult to obtain the optimal solution $\{ f_{k+m,w}^* \}$ analytically. To make it tractable, we first convert problem $\eqref{tradeoff1}$ into an equivalent LP problem via variable substitution.

\subsection{LP Problem Formulation}

To formulate an LP problem, we introduce a set of new variables $\{y_{k,w} | k \in \mathbb{K}, w \in \mathbb{W} \}$ as
\begin{align}\label{eq13}
\vspace{-0.0cm}
y_{k,w} = \sum\limits_{m=0}^{M} \pi_{k+1-m}  \theta_{m} f_{(k+1-m)+m,w}  = \sum\limits_{m=0}^{M} \pi_{k+1-m}  \theta_{m} f_{k+1,w}.\ \
\vspace{-0.0cm}
\end{align}
In Eq. \eqref{eq13}\footnote{We assume the steady-state probability whose subscript is negative is zero for notation convenience. Otherwise, variable $y_{k,w}$ should be defined as $y_{k,w} = \sum\nolimits_{m=0}^{min\{M,(k\!+\!1)\}} \pi_{k\!+\!1\!-\!m}  \theta_{m} f_{k+1,w}$.}, $\pi_{k+1-m} \theta_{m} f_{k+1,w}$ is the probability of transmitting one packet, i.e., $s[n]=1$, when there are $q[n-1]=k\!+\!1-\!m$ data packets in the buffer and $a[n]=m$ data packets newly arriving at the transmitter. Thus, $y_{k,w}$ is the probability that there are $k$ packets backlogged in the queue after one packet transmission over channel state $w$. This procedure allows us to express the objective function and the constraints of \eqref{tradeoff1} as linear functions of $\{y_{k,m}\}$. Hence, we are able to  convert the non-linear problem \eqref{tradeoff1} into a more tractable LP problem, as shown below.
\begin{theorem}\label{theorem3}
	Let $\xi = \sum_{m=1}^{M-1} \frac{m(m+1)}{2}\theta_{m+1}$ be a constant. The optimization problem \eqref{tradeoff1} is equivalent to the following LP problem:
	\begin{equation}\label{tradeoff2}
	\begin{split}
	&\mathop{\text{min}}\limits_{\{ y_{k,w} \}} \  D=\frac{1}{\bar{a}^2} \Big( \sum\limits_{k=0}^{K}  \sum\limits_{w=1}^W k \eta_w y_{k,w} - \xi \Big) \\
	& \!\!\!\! \text{s.t.}\!
	\left\{\!\!
	\begin{array}{ll}
	P=\sum\limits_{k=0}^{K}\sum\limits_{w=1}^W \eta_w P_w y_{k,w}  \leqslant P_{aver}  \!\! & (a)\\
	\sum\limits_{k=0}^{K}\sum\limits_{w=1}^W \eta_w y_{k,w} = \bar{a}    \!\!  &   (b)\\
	0 \! \leqslant \! y_{k,w} \!\! \leqslant \! \sum\limits_{m=0}^{M}  \! \theta_{m} \! \sum\limits_{i=0}^{K} \! \sum\limits_{j=1}^{W} \! G_{(k+2-m,iW+j)}\! \cdot \! y_{i,j}    \!\! & (c)\\
	\end{array}
	\right.
	\end{split}
	\end{equation}
	where $G_{(i,j)}$ is the $(i,j)$-th element of $(K\!+\!1) \!\times\! \big[W(K\!+\!1)\big]$ matrix $\bm{G}$ which describes the relationship between the steady-state probabilities $\{\pi_k\}$ of the Markov chain and the variables $\{y_{k,w}\}$, as given by
	\begin{align}\label{pi-y}
	\pi_k = \sum\limits_{i=0}^{K}\sum\limits_{j=1}^{W}  G_{(k+1,iW+j)} \cdot y_{i,j},
	\end{align}
\end{theorem}

\begin{IEEEproof}
	The detail is given in Appendix \ref{appentheorem3}.
\end{IEEEproof}

As shown in problem \eqref{tradeoff2}, there exists a minimum queueing delay for any feasible power constraint $P_{aver}$. Hence, the optimal queueing delay $D^*$ can be expressed as a function of $P_{aver}$, i.e., $D^*=d(P_{aver})$. In the following theorem, we reveal the decreasing  property of the delay-power function to discuss the structure of the optimal scheduling policy in the next section. 
\begin{theorem}\label{8171101}
	The delay function $D^{\ast}=d({P}_{aver})$ monotonically decreases with the maximum transmission power $P_{aver}$.
\end{theorem}
\begin{IEEEproof}
	The detail is given in Appendix \ref{appA}.
\end{IEEEproof}
Till now, we construct an LP problem to describe the delay-minimal scheduling problem under power constraint. After deriving the optimal solution $y_{k,w}^*$, we can then obtain the steady-state probability $\pi_k^*$ by Eq. \eqref{pi-y} and the optimal scheduling probability $\{f_{k,w}^*\}$ by Eq. \eqref{eq13}. In the sequel, we show how to derive the optimal solution as well as the optimal probabilities. 

\vspace{-0.0cm}

\section{Dual-threshold-based Policy} \label{sec5}

\begin{figure}[t]
	\centering
	\includegraphics[width=0.55\columnwidth]{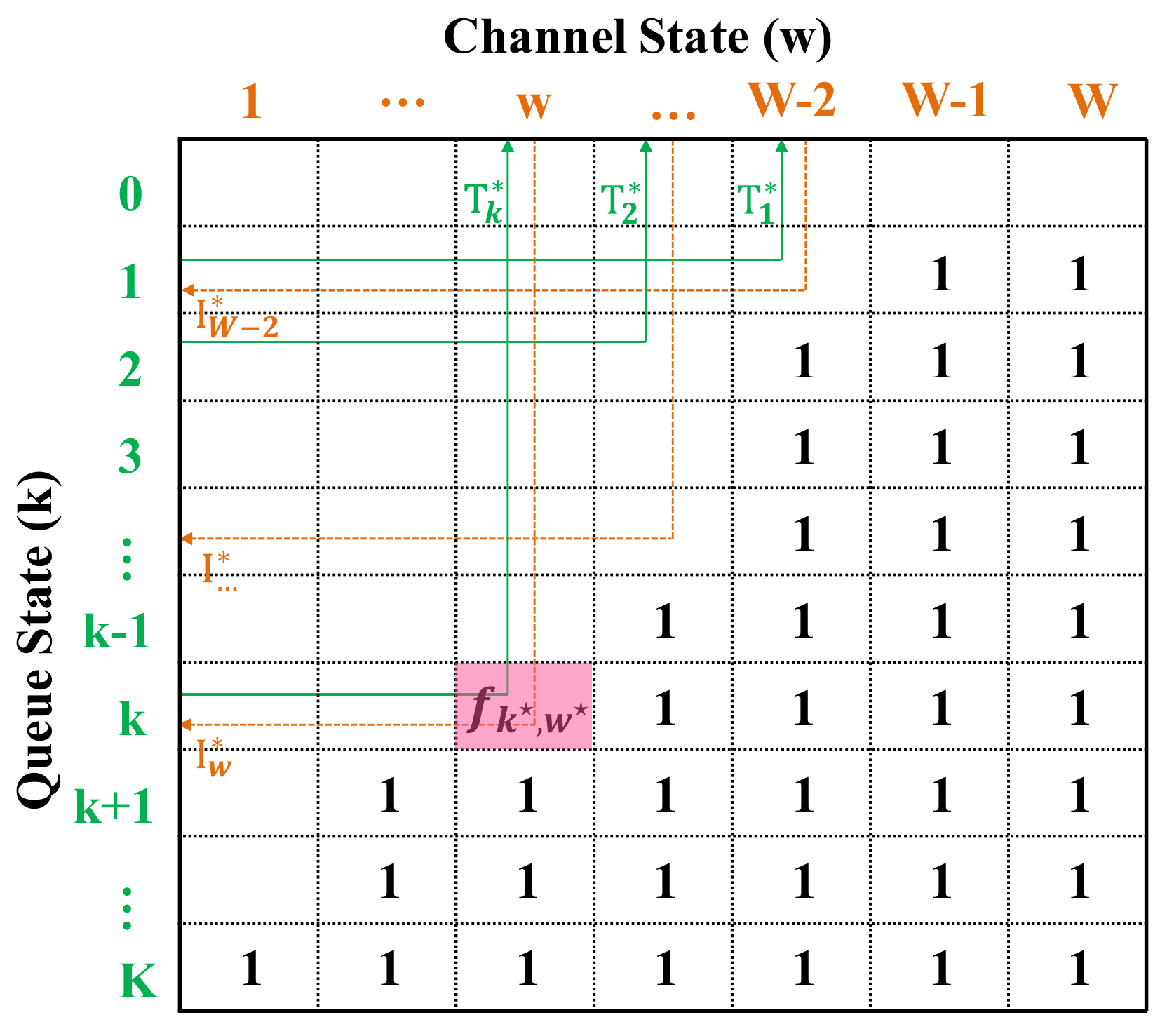}
	\caption{The dual-threshold structure: 1) for any queue length $k$, the scheduling probabilities $\{f_{k,w}\}$ follows a threshold-based structure on the channel state dimension with $T_k^*$ being the optimal threshold; 2) given the channel state $w$, the scheduling probabilities $\{f_{k,w}\}$ has a threshold-based structure on the queue state dimension with the optimal threshold $I_w^*$; 3) there is at most one threshold state at which the optimal scheduling probability is non-zero.}
	\vspace{-1cm}
	\label{6292354}
\end{figure}

In this section, we focus on revealing the dual-threshold-based structure of the optimal scheduling policy. We first present the definition of the threshold-based structure.
\begin{definition}\label{def0}
	Let $\mathbb{I}=\{0,1,2,\cdots\}$ denote an integer set. A probability set $\big\{\Upsilon_i |~i \in \mathbb{I} \big\}$ has a \textbf{$i^*$-threshold-based structure} if and only if there exists an optimal threshold  $i^*\in{\mathbb{I}}$ such that $\Upsilon_i = 0,~i < i^*$ and $\Upsilon_i = 1,~i > i^*$.
\end{definition}
In what follows, we show that the optimal scheduling policy has such a structure on both the buffer state dimension and the channel state dimension, referred to as a dual-threshold-based policy. An example of the structure is illustrated in \figurename\ref{6292354}, where positive scheduling probabilities with the indexes of buffer and channel states are plotted, and zero scheduling probabilities are omitted for briefness. In particular, given the queue state $k$, the optimal scheduling probabilities $\{f_{k,w}^*\}$ follows a threshold-based structure, i.e., $f_{k,w}^*=1$ for $w>T_k^*$ and $f_{k,w}^*=0$ for $w<T_k^*$, where $T_k^*$ is the optimal threshold on the channel state dimension. Similarly, given the channel state $w$, the optimal scheduling probabilities $\{f_{k,w}^*\}$ has a threshold-based structure on the queue state dimension. That is, there exists an optimal threshold $I_w^*$  on the queue state such that $f_{k,w}^*=1$ for $k>I_w^*$ and $f_{k,w}^*=0$ for $k<I_w^*$, respectively.
The proof of the dual-threshold-based policy is presented in two steps in subsections A and B, in accordance with the two dimensions of the channel and buffer states. What's more, we show that there is at most one threshold state ($k^\star, w^\star$) at which the optimal scheduling probability is non-zero in subsection C. Simplified threshold policy is proposed to achieve suboptimal performance in subsection D.


\subsection{Threshold-based Structure on the Channel State Dimension}\label{11291817}

We firstly reveal the non-decreasing property of the optimal solution $\{y_{k,w}^*\}$ to problem \eqref{tradeoff2}. Then, we equivalently transform problem \eqref{tradeoff2} into a new problem, which facilitates us to prove that $\{y_{k,w}^*\}$ has a $T_k$-threshold-based structure. By mapping $\{y_{k,w}^*\}$ back to $\{f_{k,w}^*\}$, the optimal scheduling policy is shown to have a threshold-based structure.

\begin{lemma}\label{6291113}
	The optimal solution to problem (\ref{tradeoff2}) $\{y_{k,w}^*\}$ has the following property, for any queue length $k$, 
	\begin{align}\label{629947}
	y_{k,w_1}^*\leqslant y_{k,w_2}^*, \quad \forall \quad 0 < w_1 < w_2 \leqslant W.
	\end{align}
\end{lemma}
\begin{IEEEproof}
	The detail is given in Appendix \ref{proof6291113}.
\end{IEEEproof}

Recall that, $y_{k,w}$ is the probability that there are $k$ packets left in the queue after one packet transmission over channel $w$. Thus, the physical meaning of \emph{Lemma} 2 is that, it reveals the tendency of exploiting a better channel state when one transmission has to be performed for the optimal policy.

\begin{lemma}\label{6291444}
	The LP Problem \eqref{tradeoff2} is equivalent to the following problem
	\begin{equation}\label{tradeoff3}
	\begin{split}
	&\mathop{\text{min}}\limits_{\{y_{k,w}\}} \  D=\frac{1}{\bar{a}} \Big( \sum\limits_{k=0}^{K}  k \cdot \mathop{\text{max}}\limits_{w} \{y_{k,w}\} - p_0\pi_0 - \varsigma \Big) \\
	&\text{s.t.}\!
	\left\{\!\!
	\begin{array}{ll}
	\mathop{\text{max}}\limits_{w} \{y_{k,w}\} = \sum\limits_{m=0}^{M} \theta_{m} \pi_{k\!+\!1\!-\!m}& (a)\\
	(\ref{tradeoff2}.a) - (\ref{tradeoff2}.c),\\
	\end{array}
	\right.
	\end{split}
	\end{equation}
	where $\varsigma=\sum\limits_{i=1}^{M-1} i \theta_{i+1} - \theta_0$ is a constant. 
\end{lemma}

\begin{IEEEproof}
	The detail is given in Appendix \ref{proof6291444}.
\end{IEEEproof}

With above two lemmas, we derive the threshold structure imposed on the channel state for a given queue length of the optimal solution $\{y_{k,w}^*\}$ to problem \eqref{tradeoff2} as follows:
\begin{theorem}\label{6291439}
	For any queue length $k$, there exists an optimal integer threshold $T_k^* \in \mathbb{W}$ such that the variables $\{y_{k,w}^*\}$ has a $T_k^*$-threshold-based structure, i.e.,
	\begin{align}\label{6291526}
	\left\{
	\begin{array}{ll}
	y_{k,w}^* = 0, & 0 < w < T_k^*; \\
	0 \leqslant y_{k,w}^* \leqslant  \sum\limits_{m=0}^{M}\theta_m\pi_{k+1-m}^*, &  w = T_k^*; \\
	y_{k,w}^* = \sum\limits_{m=0}^{M}\theta_m\pi_{k+1-m}^* , & w > T_k^*. \\
	\end{array}
	\right.
	\end{align}
\end{theorem}

\begin{IEEEproof}
	The detail is given in Appendix \ref{proof6291439}.
\end{IEEEproof}

On one hand, \emph{Theorem} \ref{6291439} is a stronger conclusion compared to Lemma 2 where the tendency of the optimal policy is revealed. It illustrates that one packet can only be transmitted if the channel state is better than a threshold. On the other hand, with the bond between  $\{y_{k,w}^*\}$ and $\{f_{k,w}^*\}$, the threshold structure in \emph{Theorem} \ref{6291439} reflects the structure of the optimal scheduling policy $\{f_{k,w}^*\}$. Specifically, the optimal scheduling probability $f_{k,w}^*$ is derived according to Eq. \eqref{eq13} and given as 1) $f_{k,w}^* = 0$, if $w<T_k$; 2) $f_{k,w}^* = 1$, if $w>T_k$; 3) $f_{k,T_k}^* = y_{k-1,T_k}^*\Big(\sum\limits_{m=0}^{M}\pi_{k+1-m}^*\theta_m\Big)^{-1}$. Thus,  $\{f_{k,w}^*\}$ also satisfies \emph{Definition} \ref{def0} and the optimal scheduling policy has a threshold-based structure on the channel state dimension for any given queue length $k$. 

\subsection{Threshold-based Structure on the Queue Length Dimension}\label{11291824}

It is not a trivial work to reveal the threshold structure on the queue state dimension straightforwardly due to the highly complicated relationship between the variables $\{y_{k,w}^*\}$. Thus, we turn to the scheduling action $s[n]$ taken by the optimal policy. Then, we map the transmission action $s[n]$ back to the scheduling probability $\{y_{k,w}^*\}$ and find that the optimal policy also has an $I_w^*$-threshold-based structure on the queue state dimension.

\begin{lemma}\label{for_w}
	For a given channel state $w$, there exists an optimal integer threshold $I_w^* \in \mathbb{K}$ such that the optimal transmission action $s^*[n]$ has the $I_w^*$-threshold structure, namely
	\begin{align}\label{12241719}
	s^*[n]=\left\{
	\begin{array}{ll}
	0, & t[n]<I_w^*;\\
	1, & t[n]>I_w^*,
	\end{array}
	\right.
	\end{align}
	where $t[n]$ $=q[n-1]+a[n]$ denotes the updated queue state after one new packet arrival in the $n$th time slot.
\end{lemma}
\begin{IEEEproof}
	The detail is given in Appendix \ref{prooffor_w}.
\end{IEEEproof}

In \emph{Lemma} \ref{for_w}, we show that the optimal transmission action $s^*[n]$ is determined based on the updated queue state $t[n]$ and the optimal threshold $I_w^*$. Together with Eq. \eqref{eq5}, we can connect $s[n]$ to the scheduling probability $f_{k,w}^*$, and reveal that the probabilities $\{f_{k,w}^*\}$ also depend on the updated queue state $t[n]=k$ and the optimal threshold $I_w^*$: 1) $f_{k,w}^* = 0$, if $k < I_w$; 2) $f_{k,w}^* = 1$, if $k > I_w$. Thus, the optimal policy is proved to has a threshold-based structure on the queue length for any given channel state.

\subsection{Dual-threshold-based Policy}

The optimal scheduling policy turns out to be a dual-threshold-based policy, as illustrated in \figurename\ref{6292354}. We complete it as follows by specifying the values on the threshold points.
\begin{theorem}\label{6292306}
	(1) The optimal scheduling policy corresponds to a dual-threshold policy. In detail, a) for any queue length $k$, there exists a threshold $T_k^* \in \mathbb{W}$, $f_{k,w}^* = 0$ for $w < T_k^*$ and $f_{k,w}^* = 1$ for $w > T_k^*$; b) there exists $T_1^* \geqslant T_2^* \geqslant \cdots \geqslant T_K^*$. (2) There is at most one threshold state ($k^\star, w^\star$) at which the optimal scheduling probability is non-zero.
\end{theorem}
\begin{IEEEproof}
Conclusion (1-a) is exactly the threshold structure obtained in subsection A. Combining the threshold structure imposed on the queue length for a given channel state, we obtain conclusion (1-b) which describes the non-increasing property of $\{T_k^*\}$.  The proof of conclusion (2) is given in Appendix G. 
\end{IEEEproof}

According to our proposed scheduling scheme, once the queue length exceeds $\max\limits_w \{I_w^*\}$, one packet will be transmitted whatever the channel state. Thus, if we set the buffer capacity $K>\max\limits_w \{I_w^*\}$, the queueing length will never reach the capacity and no packet overflow will occur. The threshold structure is a tradeoff result of reducing the queueing delay and saving power resource. An intuition explanation that explains why the policy has such a structure can be found in Appendix H. 


\subsection{The Suboptimal Policy}


It is not a trivial work to obtain closed-form expressions of the thresholds even if we have revealed their properties in \emph{Theorem} 5. By solving the LP problem, we surely can obtain optimal thresholds and the non-zero scheduling parameter that might exist at one of the joint threshold points. Otherwise, we have to resort to some search methods to find these optimal thresholds directly. In what follows, we come up with two search methods in two different scenarios.


\renewcommand{\algorithmicrequire}{ \textbf{Input:}}
\renewcommand{\algorithmicensure}{ \textbf{Output:}}
\begin{spacing}{1.2}
\begin{algorithm}[t]
	\caption{An algorithm to find the suboptimal scheduling policy}  
	\begin{algorithmic}[1]  
		\REQUIRE ~~\\  
		The average power constraint: $P_{aver}$;\\  
		The dimension of the channel state: $W$;\\
		The buffer capacity: $K$;\\ 
		The table that with a sufficiently large capacity: $Table$;\\  
		\ENSURE ~~\\ 
		The threshold on the queue states: $k^\circ$;\\
		The threshold on the channel states for $(0,k^\circ]$: $w_1^\circ$;\\
		The threshold on the channel states for $(k^\circ,K]$: $w_2^\circ$;
		\IF {$Table == NULL$}   
		\STATE initialize $Table[K][W][W][2]=\infty$; \ \ \  \# build up the table that stores the delay and power for all the deterministic policies
		\FOR{each $k^\diamond \in \{1, 2, \cdots, K\}$}  
		\FOR{each $w_1 \in \{1, 2, \cdots, W\}$}
		\FOR{each $w_2 \in \{1, 2, \cdots, W\}$}
		\STATE Set the scheduling parameters as:
		\IF {$k \leqslant k^\diamond$}		
		\STATE $f_{k,w}=0$, if $w \leqslant w_1^\circ$;
		\STATE $f_{k,w}=1$, if $w > w_1^\circ$;
		\ELSE
		\STATE $f_{k,w}=0$, if $w \leqslant w_2^\circ$;
		\STATE $f_{k,w}=1$, if $w > w_2^\circ$;
		\ENDIF
		\STATE Calculate the average queueing delay based on Eq (13): $Delay$;
		\STATE Calculate the average power based on Eq (14): $Power$;
		\STATE $Table[k][w_1][w_2] = [Delay \ Power]$;
		\ENDFOR
		\ENDFOR
		\ENDFOR
		\ENDIF
		\STATE $index$ $\leftarrow$ \{$Table:Power <= P_{aver}$\}  \ \ \  \# look up the table, find the policies that consume less power than $P_{aver}$
		\STATE $delay$ $\leftarrow$ $Min\{Table[index]:Delay$\}  \ \ \  \# find the policy that generates the smallest delay
		\STATE [$k^\circ,\ w_1^\circ,\ w_2^\circ$] $\leftarrow$ $GetIndex\{Table:Delay == delay$\}  \ \ \  \# return the threshold parameters of the suboptimal policy 
	\end{algorithmic} 	 
\end{algorithm}
\end{spacing}

Scenario I: we develop a structured search algorithm to find a suboptimal solution by fully exploiting the non-increasing properties of the optimal thresholds, as presented in \emph{Theorem} 5. In other words, this property helps to reduce the search space of the candidate threshold points significantly. In detail, combing the non-increasing property of the threshold points $T_k$, i.e., $T_{k_1} \geqslant T_{k_2}$ if $k_1 \leqslant k_2$, and the fact that the buffer capacity $K$ is usually greater than the number of channel states $W$, we know some neighbor queue lengths are likely to share a same threshold $T_k$. Based on this property, we can reduce the number of the thresholds points that need to be searched. In detail, the queue length range $[0,K]$ can be divided into several small intervals, each of which is assigned one threshold imposed on the channel states. Thus, we only need to determine how to divide the queue states and assign one threshold for each small interval. The simplified suboptimal policy is given in \emph{Algorithm} 1, where the total queue states is divided into two sub-intervals. A table can be built up to store the induced delay and power metrics for all the $\frac{1}{2}KW^2$ simple policies. Then, to obtain the suboptimal policy for a given power constraint, we only need to look up the table and return the thresholds. The performance can be further improved by assigning one scheduling probability to some threshold points.

Scenario II: we find the optimal thresholds  by looking up a preset table. Specifically, we first set up a table containing the delay and power information of all possible candidates of the optimal thresholds $\{T_k^*, I_w^*\}$ by performing intensive computations. The table formulation surely costs a lot of computation resources, but it can be done once for all.  Afterwards, when designing the optimal scheduling policy given the power constraint $p_{aver}$, we can first find the optimal thresholds $\{T_k^*, I_w^*\}$ by comparing $p_{aver}$ with the power information in the table. The optimal scheduling parameter $f_{k^\star,w^\star}$ can be further determined with the obtained $\{T_k^*, I_w^*\}$ and $p_{aver}$. And different power constraints lead to different optimal thresholds and hence the optimal scheduling parameters.


\vspace{0.0cm}
\section{Numerical Results} \label{sec6}
\vspace{0.0cm}


\begin{figure*}[t] 
	\centering	
	\subfigure[Optimal delay-power tradeoff curves]{\label{f41}
		\begin{minipage}[c]{0.5\columnwidth}
			\centering
			\includegraphics[width=\columnwidth]{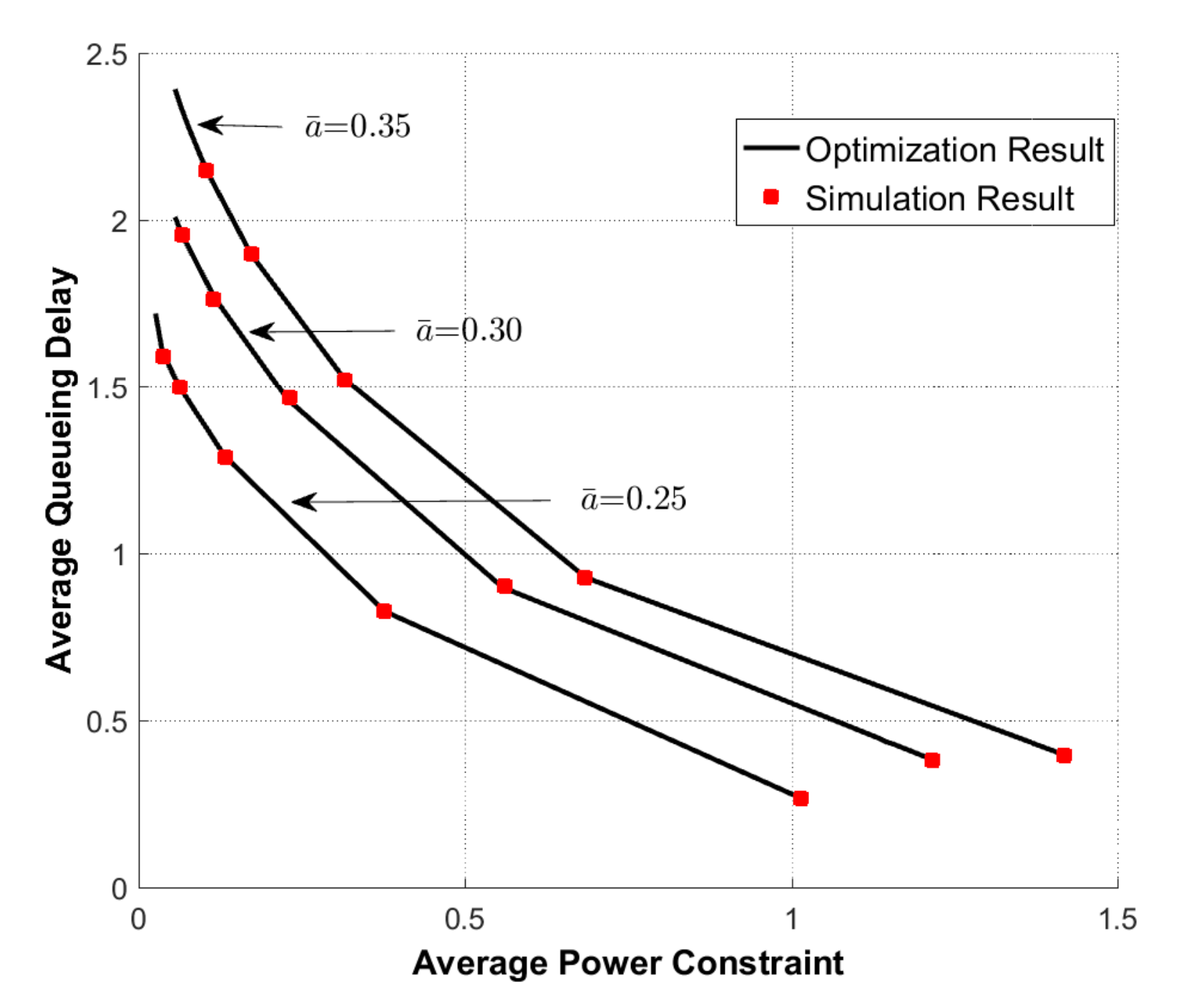}
		\end{minipage}
	}	
	\subfigure[Simulation settings]{\label{f42}
		\begin{minipage}[c]{0.3\columnwidth}
			\centering
			\includegraphics[width=\columnwidth]{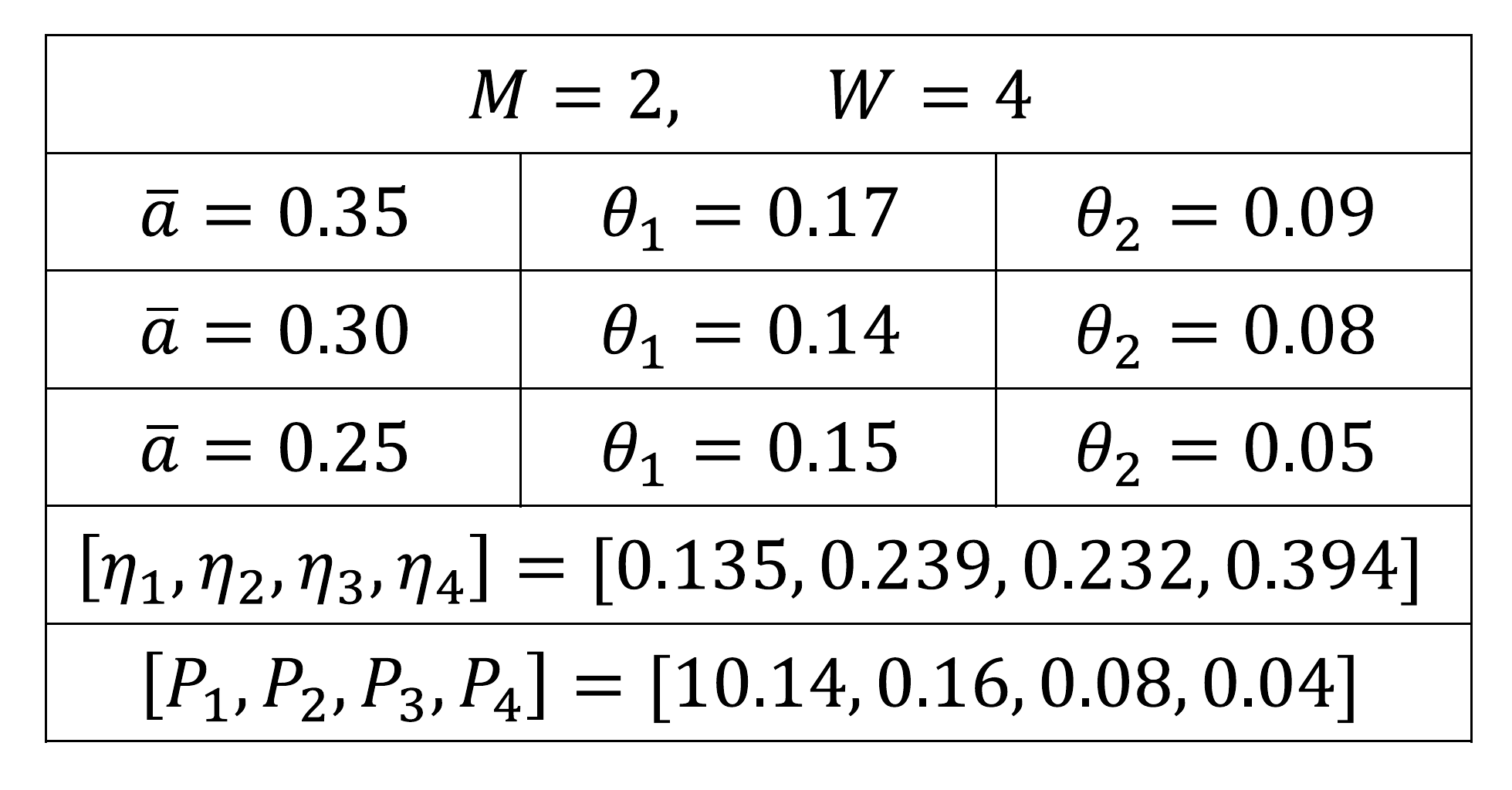}
		\end{minipage}
	}
	\caption{Optimal delay-power tradeoff curves under different arrival rates}\label{sim}
\end{figure*}
\begin{figure*}[t] 
	\centering	
	\subfigure[Optimal delay-power tradeoff curves]{\label{f51}
		\begin{minipage}[c]{0.5\columnwidth}
			\centering
			\includegraphics[width=\columnwidth]{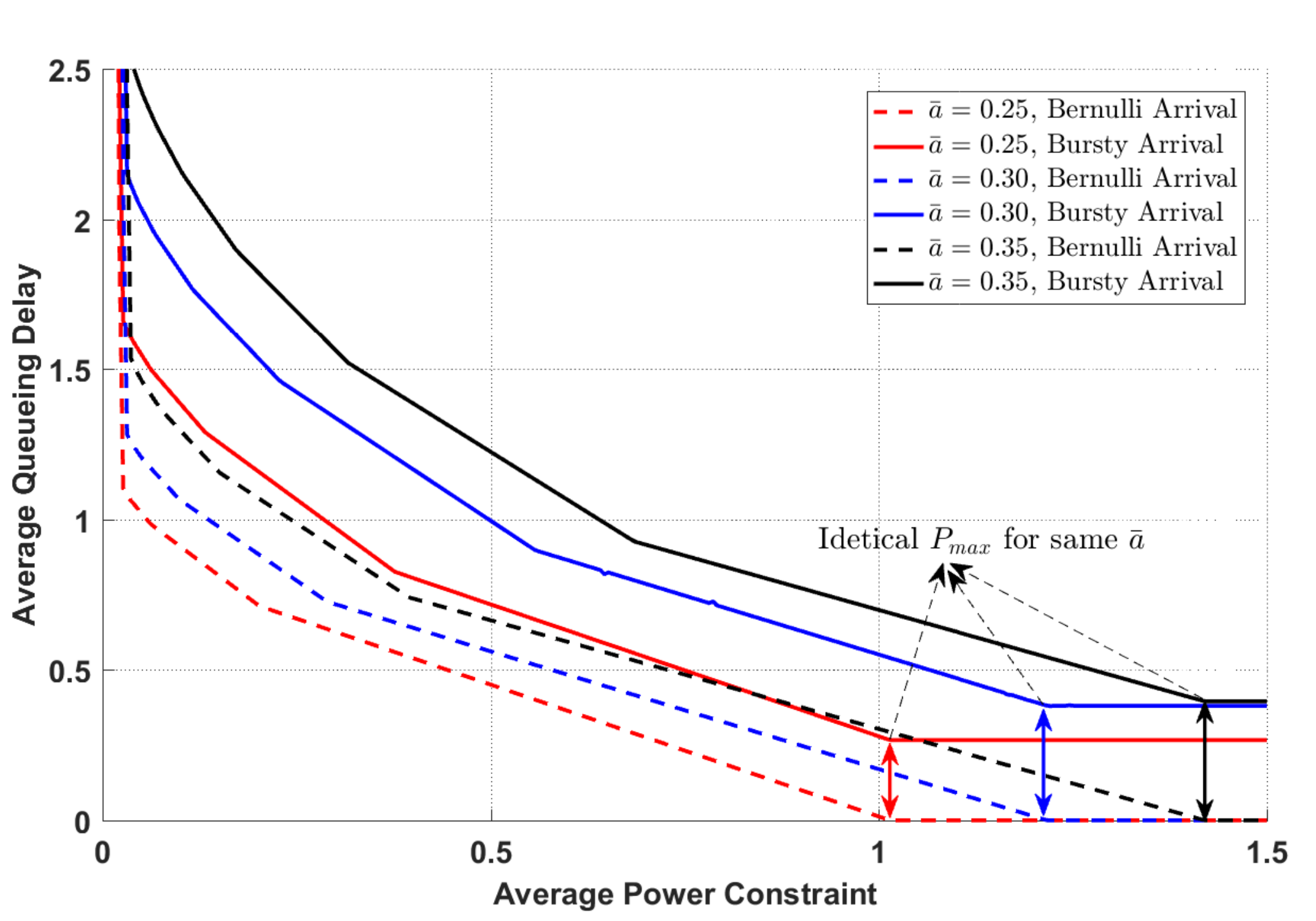}
		\end{minipage}
	}	
	\subfigure[Simulation settings]{\label{f52}
		\begin{minipage}[c]{0.3\columnwidth}
			\centering
			\includegraphics[width=\columnwidth]{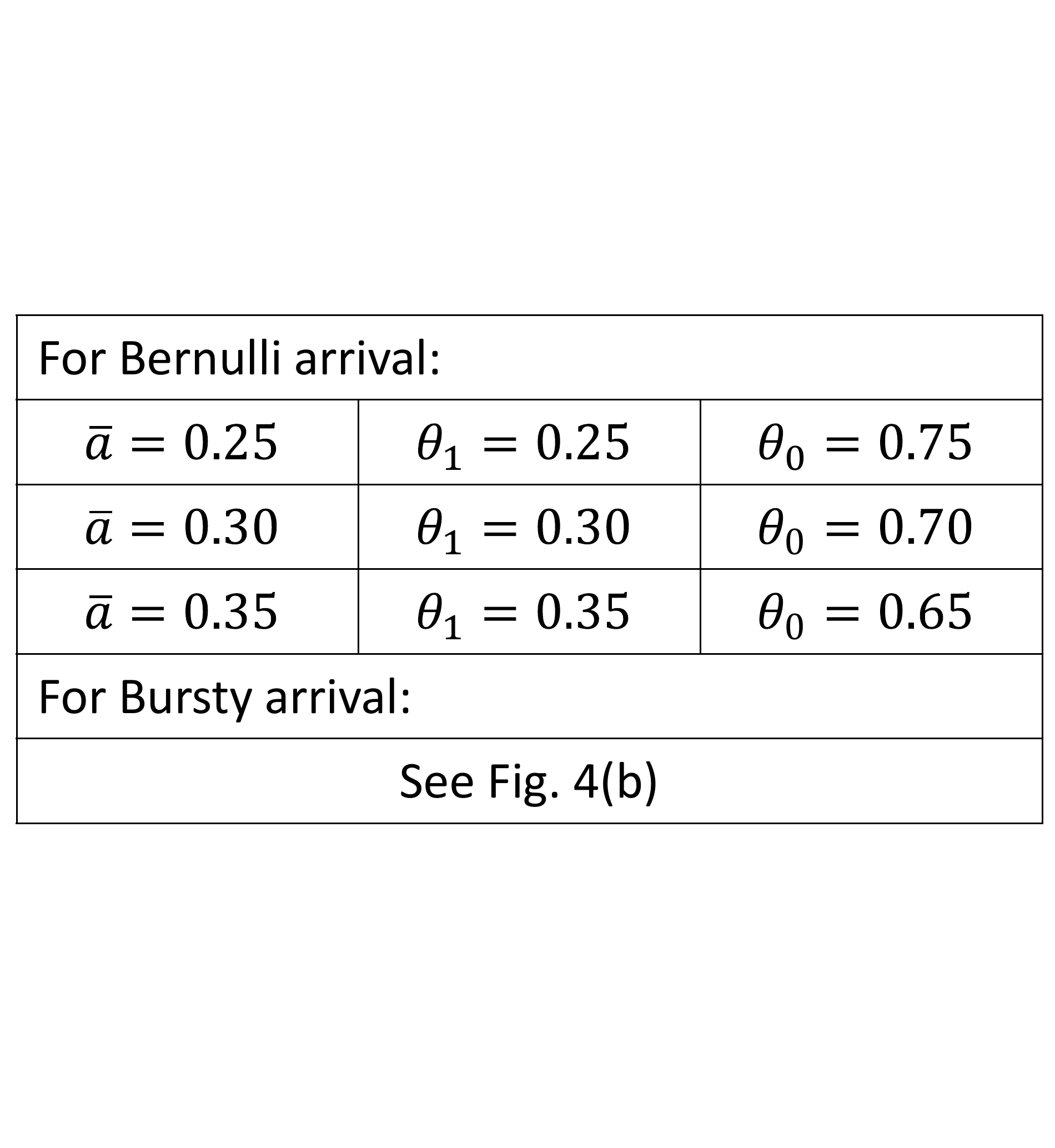}
		\end{minipage}
	}
	\caption{The effect of the burstiness of the arrival}\label{Bernoulli}
\end{figure*}
\begin{figure*}[t] 
	\centering	
	\subfigure[Optimal delay-power tradeoff curves]{\label{f61}
		\begin{minipage}[c]{0.5\columnwidth}
			\centering
			\includegraphics[width=\columnwidth]{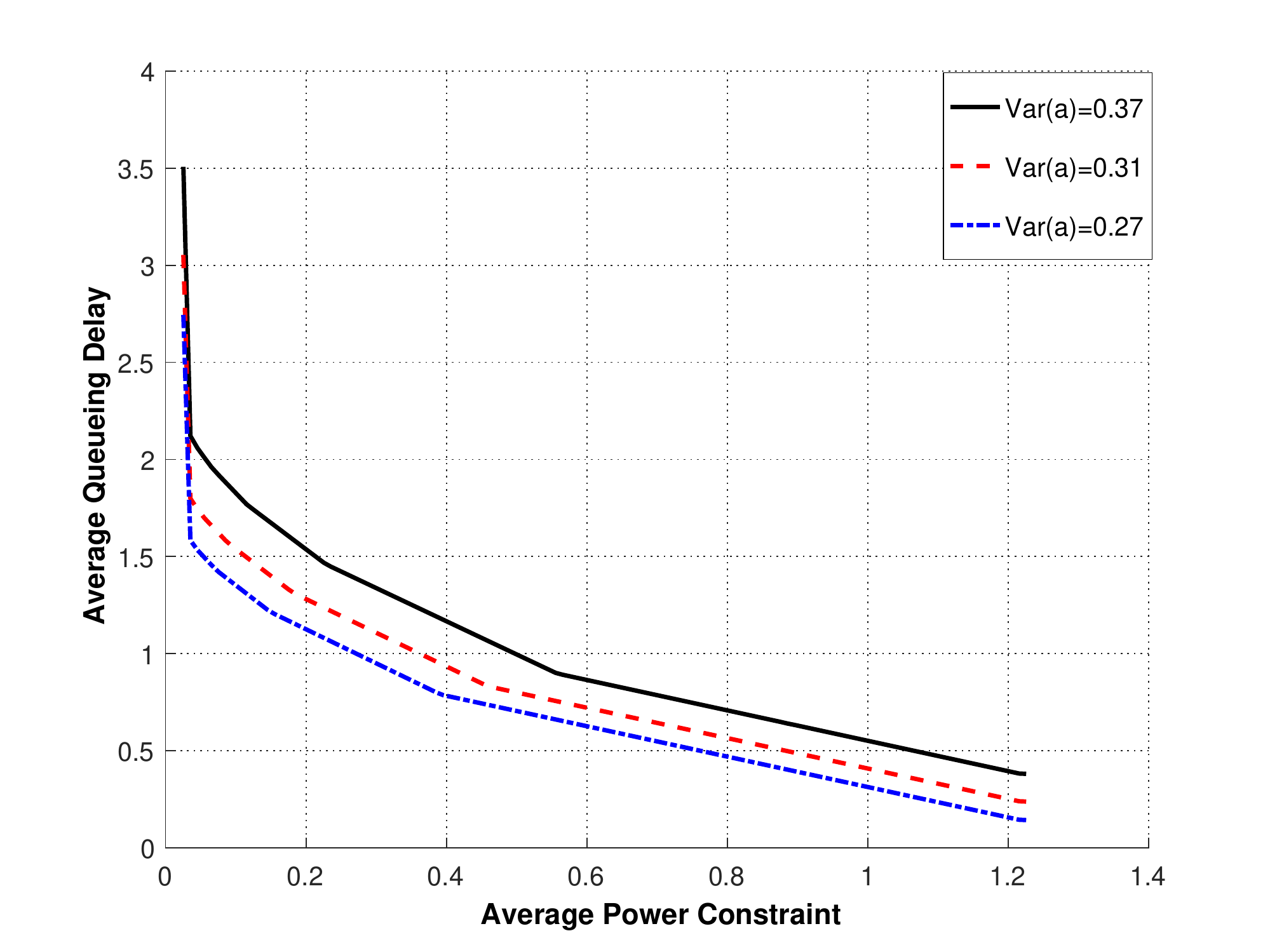}
		\end{minipage}
	}	
	\subfigure[Simulation settings]{\label{f62}
		\begin{minipage}[c]{0.3\columnwidth}
			\centering
			\includegraphics[width=\columnwidth]{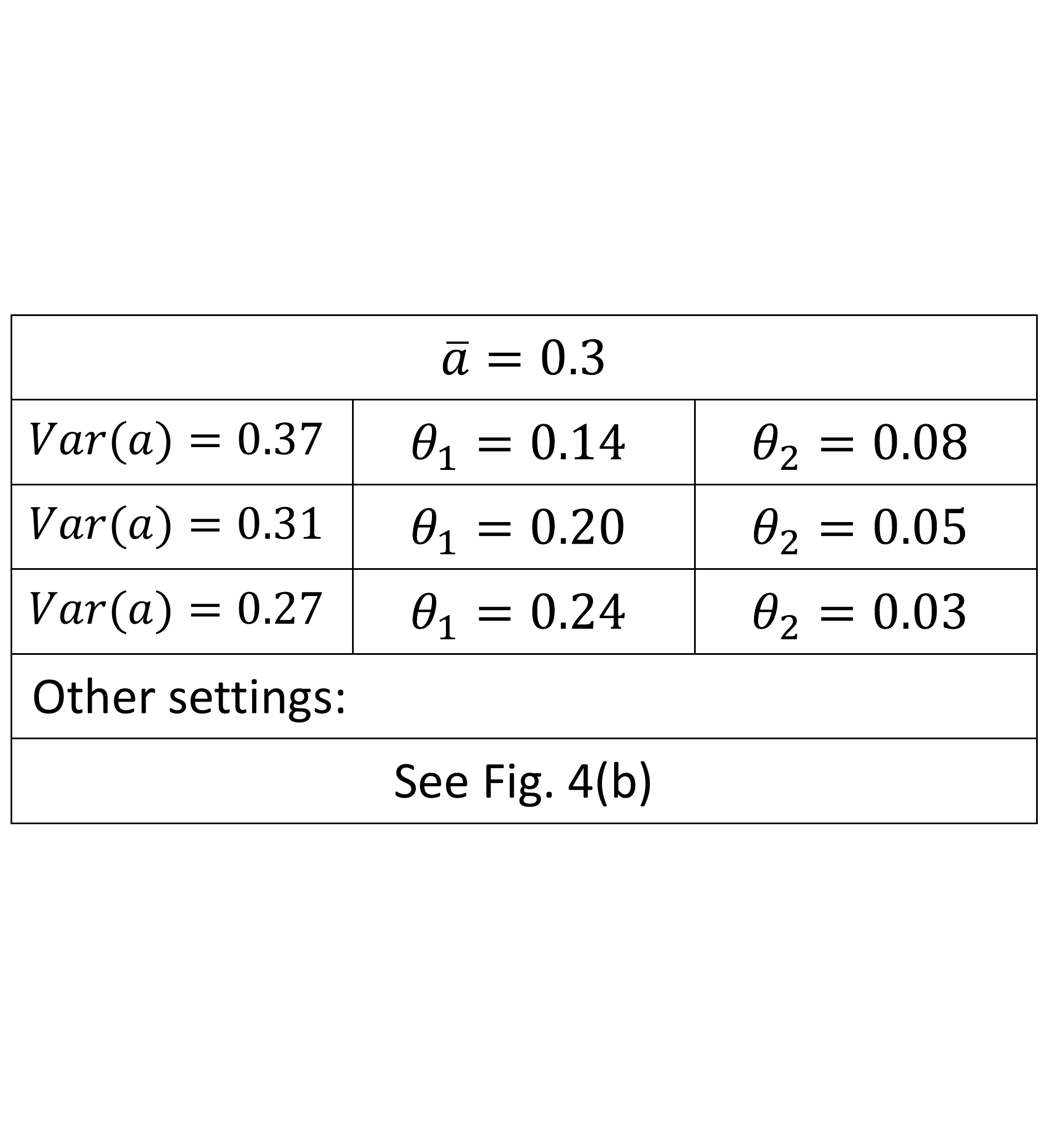}
		\end{minipage}
	}
	\caption{Optimal delay-power tradeoff curve in different arrival variances}\label{sim_2}
\end{figure*}
\begin{figure*}[htb]
	\centering
	\captionsetup[subfigure]{labelformat=simple,captionskip=6bp,nearskip=6bp,farskip=0bp,topadjust=0bp}
	\renewcommand{\thesubfigure}{(\alph{subfigure})}
	\subfigure[Power constraint $P_{aver}=0.10$,]{
		\includegraphics[width=0.3\columnwidth]{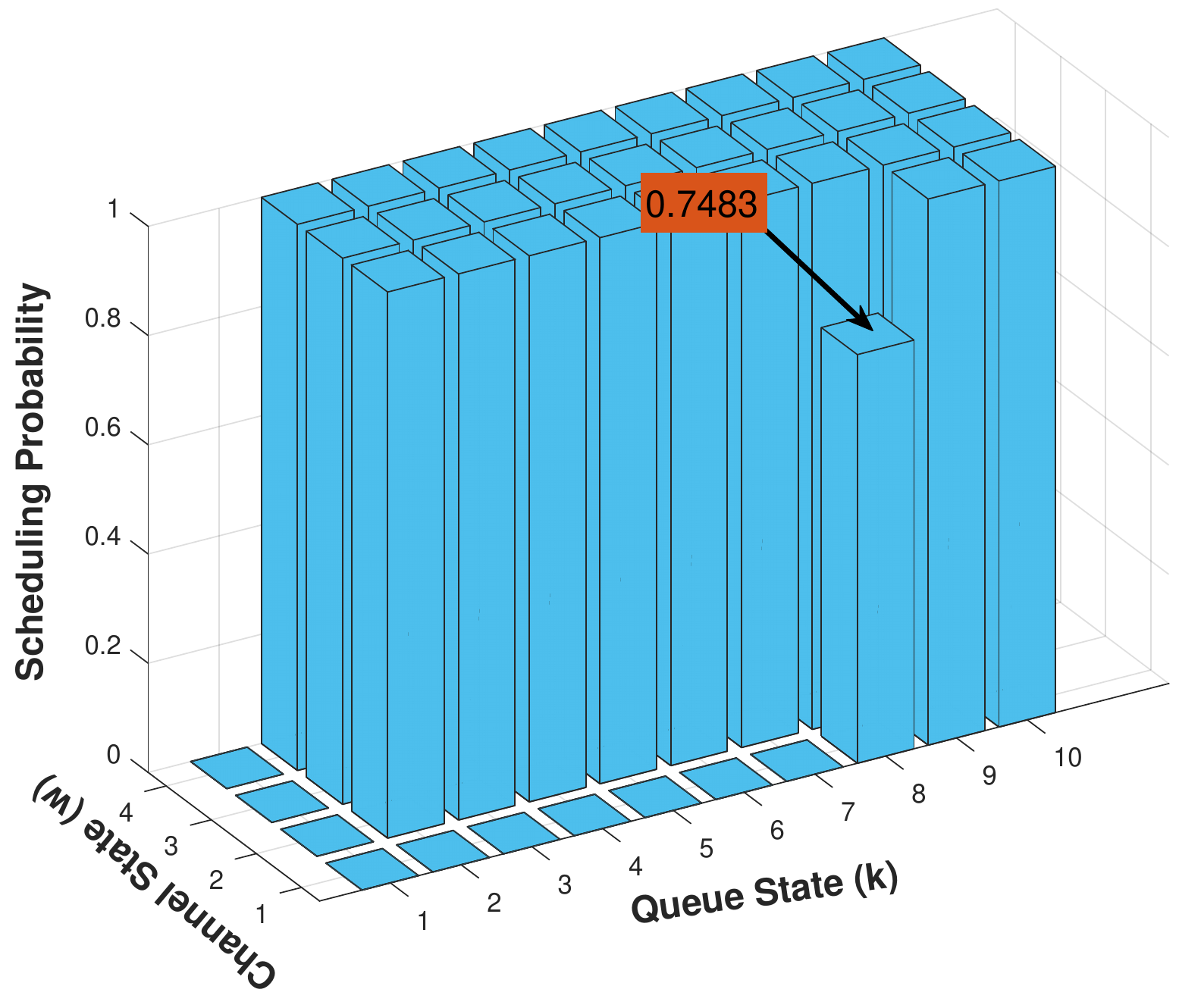} \label{3d0}}
	\hspace*{-0.3cm}
	\subfigure[Power constraint $P_{aver}=0.53$,]{
		\includegraphics[width=0.3\columnwidth]{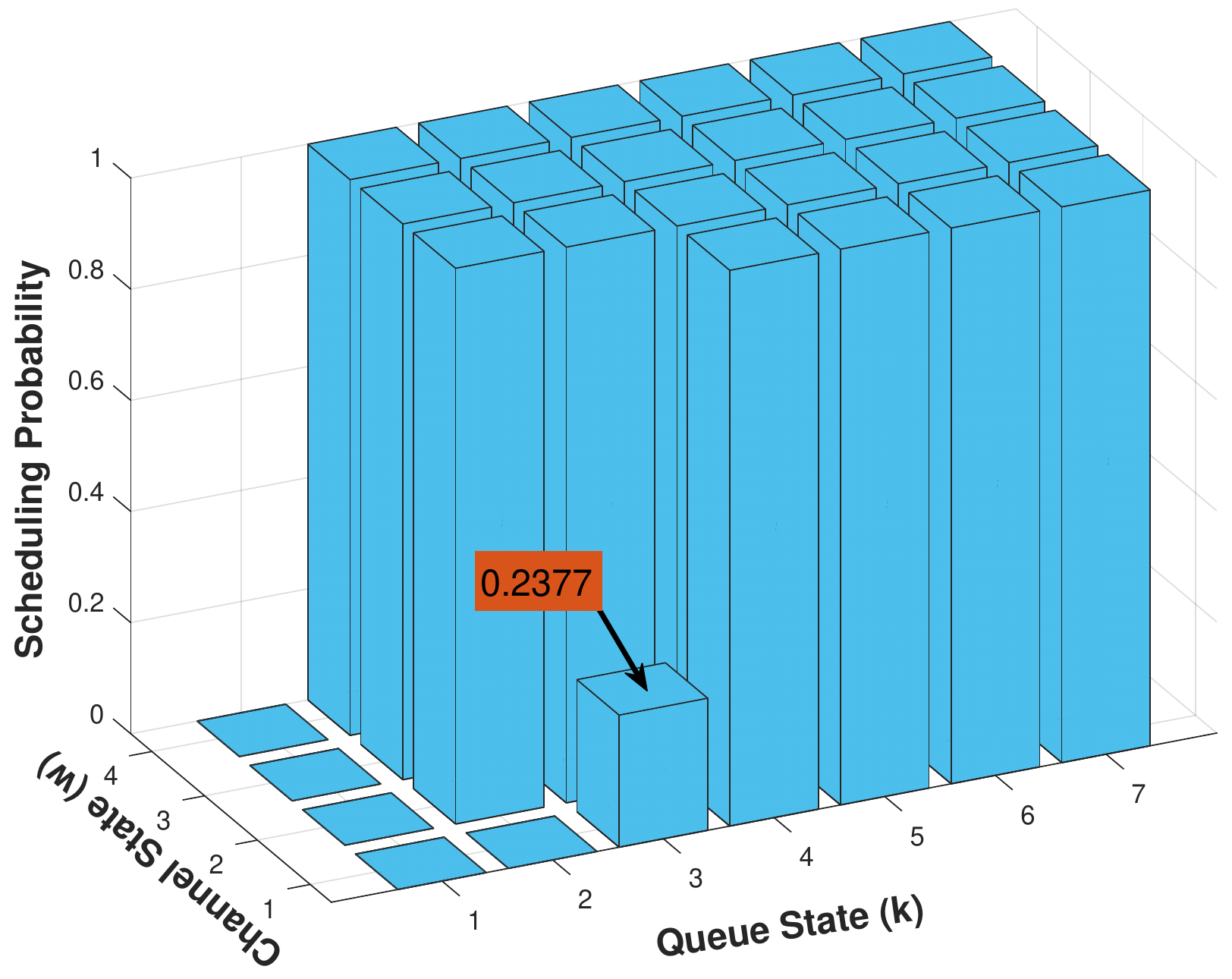} \label{3d1}}
	\hspace*{-0.3cm}
	\subfigure[Power constraint $P_{aver}=0.63$]{
		\includegraphics[width=0.3\columnwidth]{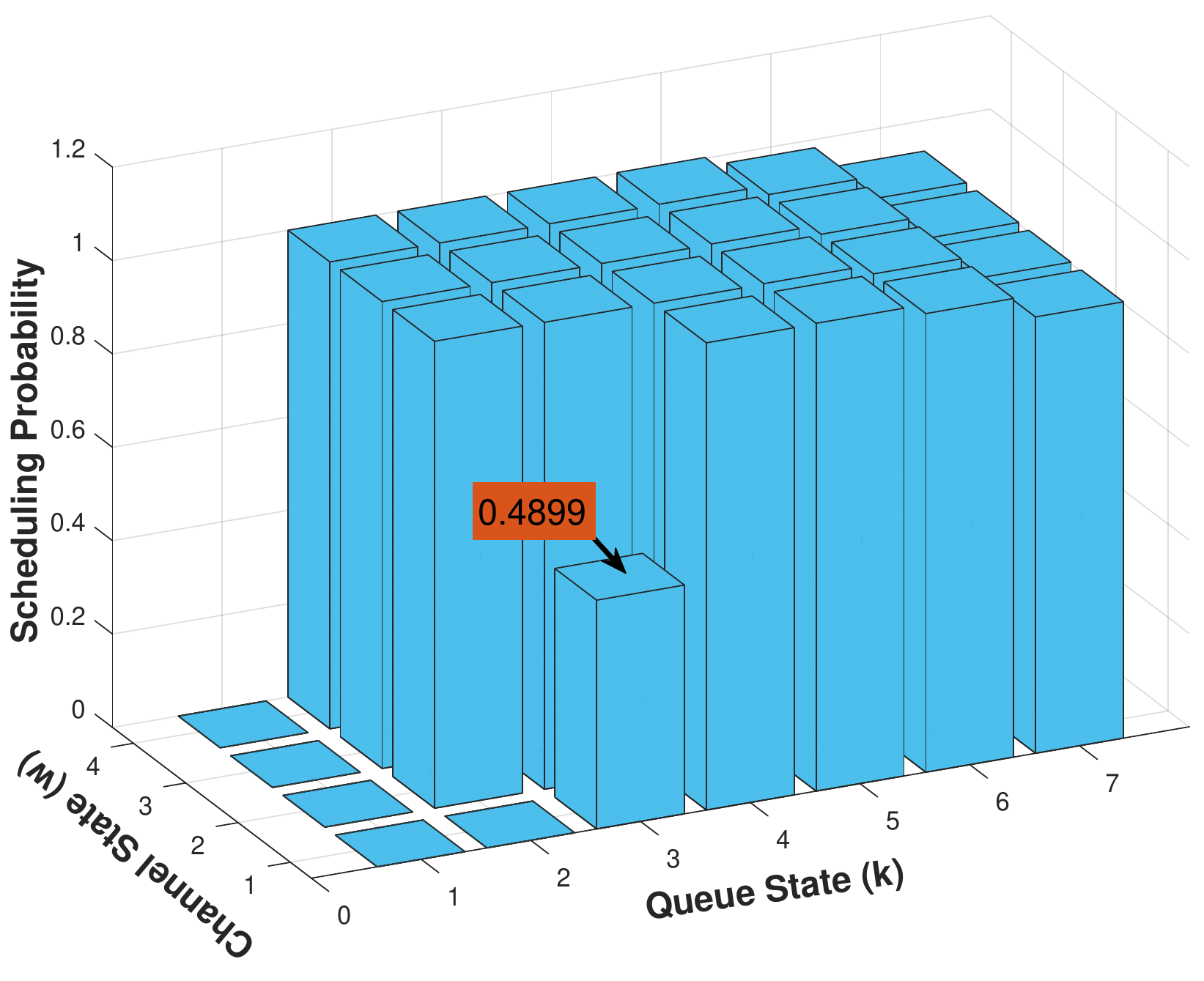} \label{3d2}}
	\caption{The dual-threshold-based policy: $\bar{a}=0.35$.}\label{3dfigure}
\end{figure*}

In this section, simulation results are given to validate the derived dual-threshold-based scheduling policy and to demonstrate its potential. For performance comparison, theoretical results of the optimal delay-power function $D^*(P_{aver})$ are obtained by solving the LP problem \eqref{tradeoff2}. Meanwhile, simulation results are obtained by applying the dual-threshold-based scheduling policy with the optimal transmission parameters. In simulations, data packets are generated following a given probabilistic distribution $\{ \theta_{m} \}$. The $W$-state block fading channel model is adopted and follows with probability $\{\eta_m\}$. Each simulation runs over $10^6$ time slots. As shown in \figurename\ref{sim}-\ref{3dfigure}, the theoretical and simulation results are plotted by lines (solid or dashed) and marked by red square dots, respectively.

\figurename\ref{sim} plots the delay-power tradeoff curves under different average packet arrival rates. The simulation results are in good agreement with the theoretical results, which validates the optimality of the derived dual-threshold-based policy. The delay-power tradeoff curve is piecewise linear since the threshold-based is obtained as the linear combinations of deterministic scheduling parameters. Besides, the average delay monotonically decreases with the maximum average power, as stated in \emph{Theorem} \ref{8171101}. When the power constraint $P_{aver}$ decreases to zero, the queueing delay increases dramatically to infinity, which implies that the queueing system is unstable. Given the same power constraint, the queueing delay increases with the packet arrival rate since more packets are detained in the buffer due to lack of transmission opportunities.

In \figurename\ref{Bernoulli}, we evaluate the effect of the burstiness of the packet arrival on the optimal delay-power tradeoff curves, considering different packet arrival patterns, namely, the Bernoulli arrival and the bursty arrival. We can see that the proposed scheduling policy has a better delay-power tradeoff performance when the packet arrivals follow the Bernoulli distribution rather than the more bursty probabilistic distribution (with larger variance), subject to the same average arrival rate. This is due to the fact that the bursty packet arrivals bring more randomness to the queueing system. The average queueing delay decreases with the increase of the power constraint and remains constant when the power constraint exceeds a constant $P_{max}$. In other words, the delay-power curve becomes flat after an inflection point $(D^*_{min}, P_{max})$,  where $D^*_{min}$ is the globally minimum delay and $P_{max}$ denotes the power consumption that the source spends to keep transmitting packets as long as the buffer is not empty, regardless of the channel state. However, the value of $P_{max}$ is identical for the two different patterns. The value of $D_{min}$ is able to reach zero for the Bernoulli arrival since the transmission rate is fixed as one packet per slot and is greater than zero due to the burstiness.   

Inspired by the observation in \figurename\ref{Bernoulli}, we further demonstrate the delay-power tradeoffs in \figurename\ref{sim_2} for the packet arrivals have the same average arrival rate and different variances. It is observed that a higher queueing delay is induced when the data arrival variance is larger. Due to higher bursty arrivals, some packets have to wait for a longer time before they are transmitted, which leads to a larger queueing delay.

In \figurename\ref{3dfigure}, we demonstrate the theoretical results to validate the dual-threshold-based  structure of the optimal scheduling policy, which are in agreement with the structure shown in \figurename\ref{6292354}. The transmission probabilities reveal a threshold-based structure on both the channel state dimension and the queue length dimension. In \figurename\ref{3d0}, the threshold is in channel state $1$ and queue length $8$ while in \figurename\ref{3d1}, it is in channel state $1$ and queue length $3$. Thus, transmission is much easier to occur in \figurename\ref{3d1}, which corresponds to a higher power consumption. That is, the scheduler makes use of the power resource mainly by adjusting the threshold point for quite different power constraints. In \figurename\ref{3d1} and \figurename\ref{3d2}, it's calculated for both scenarios that the threshold is in channel state $1$ and queue length $3$. However the scheduler makes a decision of transmitting one packets with probability $0.2377$ in \figurename\ref{3d1} and $0.4899$ in \figurename\ref{3d2} on the threshold point, respectively. That is, the scheduler makes full use of the power resource mainly by adjusting the transmission probability on the threshold point for slight different power constraints. 

\begin{figure}[t]
	\centering
	\includegraphics[width=0.6\columnwidth]{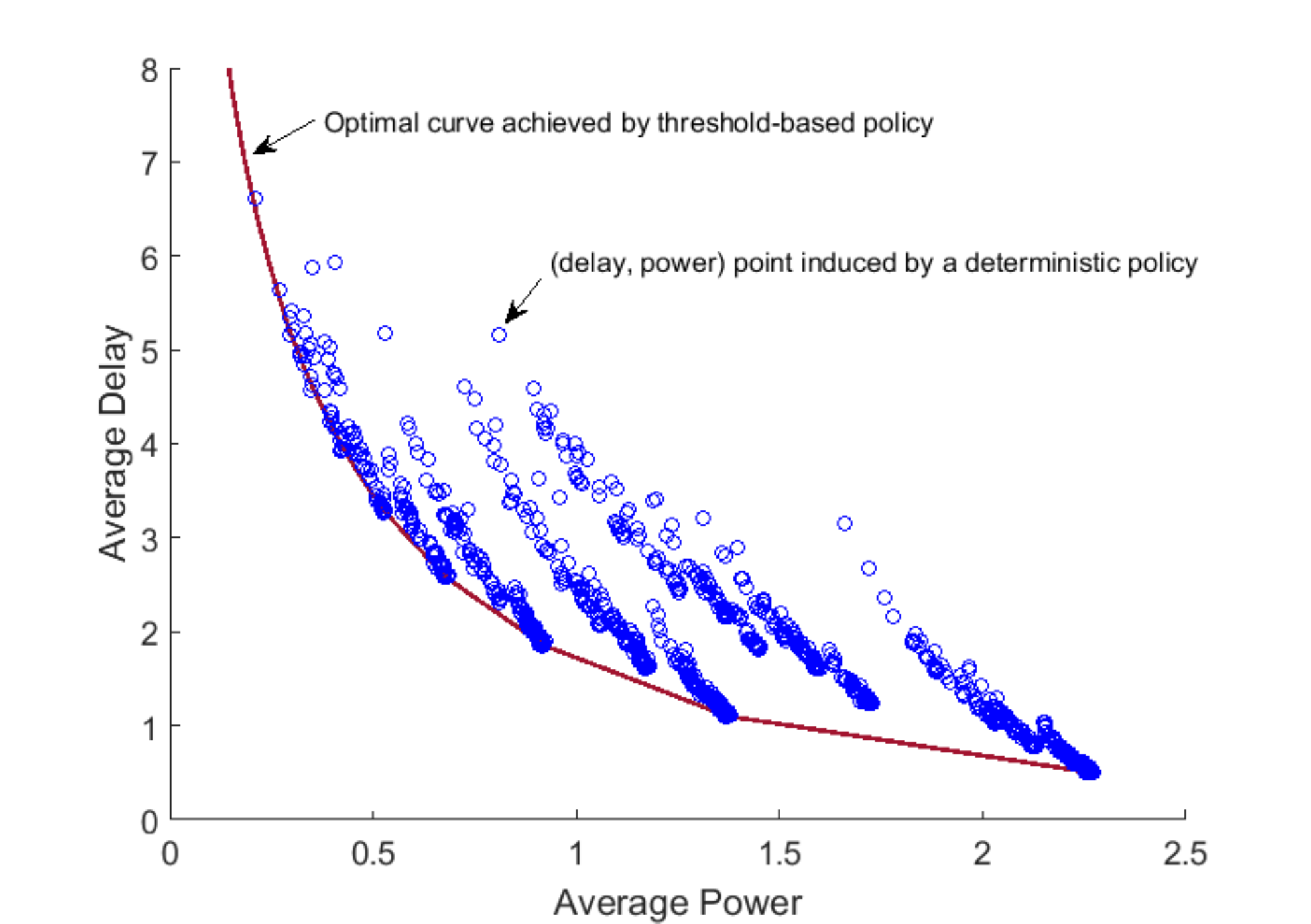}
	\caption{The optimal delay-power tradeoff is the lower boundary of the achievable $(D, P)$ region: the arrival distribution is $\bar{a}=0.55$, $\theta_1=0.3$, $\theta_2=0.125$; the channel distribution is $[\eta_1=0.6, \eta_2=0.4]$ and $[P_1, P_2]= [10.14, 0.103]$.}
	\label{compare1}
\end{figure}

In \figurename\ref{compare1}, we plot the optimal delay-power curve of our proposed scheme and 1000 delay-power points of the deterministic policy with the binary transmission parameters $f_{k,w}\in\{0,1\}$ randomly generated. As can be seen from this figure, the  delay-power tradeoff curve is the lower boundary of the convex hull of the achievable delay-power region, which is in accordance with the conclusion proved in [21] that the optimal probabilistic policy can be constructed by the convex combination of deterministic scheduling policies. Hence, our proposed optimal scheduling policy outperforms any deterministic scheduling policies given the same power constraint. Meanwhile,  our proposed stochastic scheduling policy with the optimal thresholds and scheduling parameters can achieve the same optimal delay-power tradeoff performance as the optimal scheduling policies found by the DP method.


\begin{figure}[t]
	\centering
	\includegraphics[width=0.6\columnwidth]{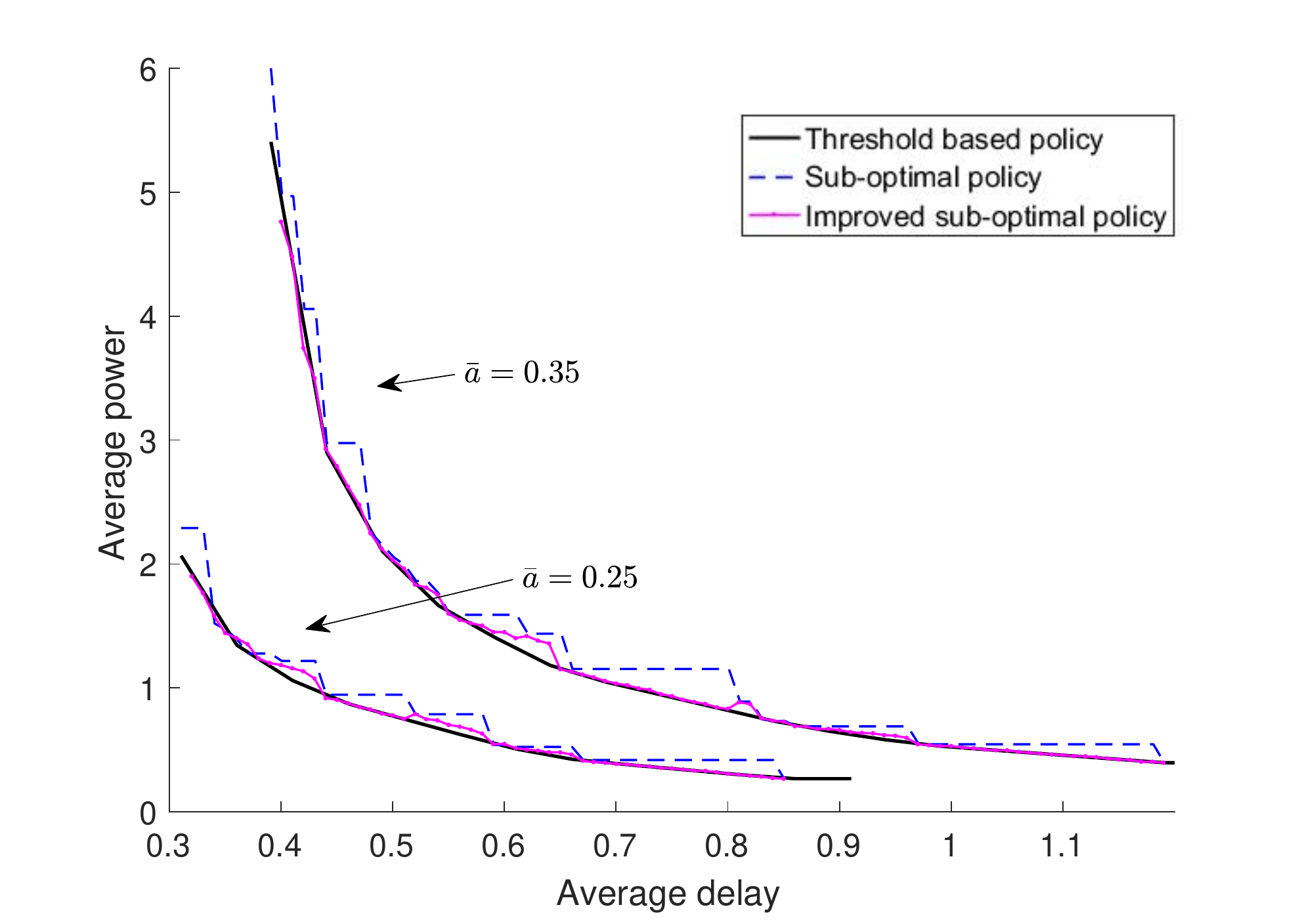}
	\caption{The tradeoff curve induced by the suboptimal policy: the arrival distribution is $\bar{a}=0.55$, $\theta_1=0.3$, $\theta_2=0.125$; the channel distribution is $[\eta_1=0.135, \eta_2=0.239, \eta_3=0.232, \eta_4=0.394, ]$ and $[P_1, P_2, P_3, P_4]= [10, 5, 2, 1]$.}
	\label{compare2}
\end{figure}

In \figurename\ref{compare2}, we plot the delay-power tradeoff curves induced by the optimal policy and suboptimal policy proposed in subsection V-D. Since same suboptimal policy will be assigned to close power constraints, the suboptimal curve remains flat sometimes. Also, the suboptimal policy may indeed be the optimal policy for some power, thus, there exist intersections for the two curves. The suboptimal policy determines the scheduling with less threshold points compared to the optimal one, thus, it is easier to be found and apply but with less accuracy. One can also assign a probability for the threshold points and further make use of the power as given in the purple curves.

\vspace{0.0cm}
\section{Conclusion} \label{sec7}
\vspace{0.0cm}

In this paper, we studied the power-constrained delay-optimal scheduling problem in wireless systems, where arbitrary packet arrivals and multi-state block-fading channels were considered. A probabilistic queue-aware and channel-aware scheduling policy was proposed to schedule packet transmissions over a $W$-state wireless fading channel and investigated in the framework of constrained MDP. Through theoretical analysis, we reveal the  dual-threshold-based structure of the optimal scheduling policy. It is found that the optimal scheduler always seeks to exploit a good channel while maintaining a relatively short queue as possible to reduce the latency. To this end, the scheduler should schedule packet transmissions based on the queue state and the channel state. Specifically, given a channel state, if the queue length exceeds the threshold, the transmitter  should transmit to decrease the latency. Otherwise, it should keep silent to save power. In the future, we will extend this work to more general  scenarios with adaptive-rate transmission and/or multi-user scheduling.
\vspace{0.0cm}

\appendices

\section{The proof of \emph{Theorem} \ref{theorem3}}\label{appentheorem3}

In this appendix, we show that problem \eqref{tradeoff1} can be equivalently converted into LP problem \eqref{tradeoff2} with variables $\{y_{k,w}\}$ being the optimization variables. To make it clear, we explain the transformation procedure in the following five steps. We first present the equivalent expressions of the average queueing delay and the power constraint in Part \ref{8162233}. Secondly, we specify the ranges of optimization variables $\{y_{k,w}\}$ corresponding to constraint (\ref{tradeoff1}.b) in Part \ref{761609}. Then,  we reformulate constraints (\ref{tradeoff1}.c) and (\ref{tradeoff1}.d) in Part \ref{11201308} and Part \ref{761610}, respectively. Finally, we explain why constraints (\ref{tradeoff1}.c) and (\ref{tradeoff1}.e) are not shown in the LP problem \eqref{tradeoff2} in Part \ref{11201529}.

\subsubsection{The average queueing delay and the power constraint can be re-expressed as
}\label{8162233}
\begin{align} \label{new_delya_power}
\left\{
\begin{array}{l}
D=\frac{1}{\bar{a}^2} \Big( \sum\limits_{k=0}^{K} \sum\limits_{w=1}^W k \eta_w y_{k,w} - \xi \Big),  \\
P=\sum\limits_{k=0}^{K}\sum\limits_{w=1}^W \eta_w P_w y_{k,w}, 
\end{array}
\right.
\end{align}
\emph{where} $\xi = \sum_{m=1}^{M-1} \frac{m(m+1)}{2}\theta_{m+1}$.

Firstly, we re-express the average queueing delay. Adding the weighted sum of the terms $\sum\nolimits_{w=1}^{W}\eta_w y_{k,w}$ with $k$ being the weight, we have
\begin{align}
&\sum\limits_{k=0}^{K} k \sum\limits_{w=1}^W \eta_w y_{k,w} =^{\textcircled{\scriptsize 1}} \sum\limits_{k=0}^{K} k \sum\limits_{m=0}^{M-1} \pi_{k-m}\sum\limits_{i=m+1}^{M}\theta_i    \nonumber \\
&=^{\textcircled{\scriptsize 2}} (\theta_1 + \theta_2 + \cdots + \theta_M) \sum\nolimits_{k=0}^{K} k \pi_k + (\theta_2 + \cdots  + \theta_M) \sum\nolimits_{k=0}^{K} k \pi_{k-1} +\cdots + \theta_M \sum\nolimits_{k=0}^{K} k \pi_{k-(M-1)}           \nonumber \\
&=^{\textcircled{\scriptsize 3}} (\theta_1 + \theta_2 + \cdots  + \theta_M) Q  + (\theta_2 + \cdots  + \theta_M) (Q+1) + \cdots  + \theta_M  (Q+(M-1))               \nonumber \\
&=^{\textcircled{\scriptsize 4}} \bar{a}Q + \sum\nolimits_{m=1}^{M-1} \sum\nolimits_{i=1}^{m} i \theta_{m+1}          \nonumber \\
&=^{\textcircled{\scriptsize 5}}\bar{a}Q + \sum\nolimits_{m=1}^{M-1} \dfrac{m(m+1)}{2} \theta_{m+1}     =    \bar{a}Q +   \xi, 
\end{align}
where equality \textcircled{\scriptsize 1} is derived by substituting \eqref{761533}, equality \textcircled{\scriptsize 2}  is obtained by expressing each term in  ${\pi_{k-m}\sum_{i=m+1}^M\theta_i}$ separately for $m=0,~ 1, ~2,\cdots, M-1$, equality \textcircled{\scriptsize 3} comes from the definition of the average queue length, equality \textcircled{\scriptsize 4} is obtained by substituting $\bar{a}=\sum_{m=0}^M m\cdot \theta_m$, and equality \textcircled{\scriptsize 5} stems from $\sum\nolimits_{i=1}^{m}i=m(m+1)/2$.
Thus, the average queue length is:
\begin{align}
Q = \frac{1}{\bar{a}} \Big( \sum\nolimits_{k=0}^{K} \sum\nolimits_{w=1}^W k \eta_w y_{k,w} - \xi \Big).
\end{align}
According to the Little's Law, we obtain the average queueing delay as given in Eq. \eqref{new_delya_power}.

Secondly, we express the average power as a function of variables $\{y_{k,m}\}$.
From the definition of variable $y_{k,w}$ in Eq. \eqref{eq13}, we have
\begin{align}\label{11172305}
\sum\limits_{k=0}^K  \pi_k  \sum\limits_{m=0}^M   \theta_m f_{k\!+\!m,w} = \sum\nolimits_{k=0}^{K} y_{k,w}.
\end{align}
By substituting Eq. \eqref{11172305} into Eq. \eqref{eq12}, we obtain the average power as presented in Eq. \eqref{new_delya_power}.

\subsubsection{	The variable $y_{k,w}$ satisfies the following inequalities}\label{761609}
\begin{align}\label{11201019}
0 \leqslant y_{k,w} \leqslant \sum\limits_{m=0}^{M } \theta_{m} \pi_{k+1-m}.
\end{align}

We know that probability $f_{k+1,w}$ takes its value from the  interval $[0,1]$. By substituting $f_{k+1,w}=0$ and $f_{k+1,w}=1$ in Eq.\eqref{eq13}, we get the lower and upper bounds of variable $y_{k,w}$, respectively. In this way, the range of variable $y_{k,w}$ is specified by the inequalities in Eq. \eqref{11201019}.



\subsubsection{The constraint (\ref{tradeoff1}.c) can be equivalently expressed as}\label{11201308}
\begin{align} \label{761533}
\sum\limits_{m=0}^{M-1} \pi_{k-m}\sum\limits_{i=m+1}^{M}\theta_i = \sum\limits_{w=1}^W \eta_w y_{k,w} = \sum\limits_{i=0}^{M-1}\pi_{k-i} r_i.
\end{align}
\emph{where} $r_i=\sum\nolimits_{m=i+1}^M ~ \theta_m$.

In fact, constraint (\ref{tradeoff1}.c) denotes the steady-state equilibrium equation of the underlying Markov chain:
\begin{align}
\pi_{k+1}\mu_{k+1}= \sum\limits_{m=0}^{M-1}\pi_{k-m} \sum\limits_{i=m+1}^{M}\lambda_{k-m,i},  \ \ k \in \mathbb{ K}.
\label{steady-equation}
\end{align}

We can obtain the following conclusion
\begin{align}\label{11201118}
\sum\limits_{i=m}^{M}\lambda_{k,i}= &\sum\limits_{i=m}^{M} \theta_i   -  \theta_m\sum\limits_{w=1}^W \eta_w f_{k+m,w}, ~~  m \in \mathbb{M}
\end{align}
by adding up the terms $\{\lambda_{k,i}~|~i=m,\cdots,M\}$ in Eq. \eqref{eq601}. 

By substituting Eq. \eqref{eq602} and Eq. \eqref{11201118} into Eq. \eqref{steady-equation}, we have 
\begin{align}
&\pi_{k+1} ( \theta_0\sum\nolimits_{w=1}^W \eta_w f_{k+1,w} )  = \sum\limits_{m=0}^{M-1 }\pi_{k-m} (\sum\limits_{i=m+1}^{M} \theta_i   -  \theta_{m+1}\sum\limits_{w=1}^W \eta_w f_{k-m+m+1,w} ).
\end{align}
Then, we can further simplify this equation as 	
\begin{align}  \label{new-varable}
\sum\limits_{m=0}^{M-1} \pi_{k-m}\sum\limits_{i=m+1}^{M}\theta_i
&= \sum_{m = -1}^{M-1} \pi_{k-m}\theta_{m+1} \sum\limits_{w=1}^{W}\eta_w f_{k+1,w} \nonumber \\
&= \sum\limits_{w=1}^{W} \eta_w ( \sum_{m = -1}^{M-1} \pi_{k-m}\theta_{m+1} f_{k+1,w} )    \\
&=\sum\limits_{w=1}^{W} \eta_w  y_{k,w}. \nonumber
\end{align}

\subsubsection{The constraint (\ref{tradeoff1}.d) can be re-expressed as $\sum\nolimits_{k=0}^{K}\sum\nolimits_{w=1}^W \eta_w y_{k,w} = \bar{a}.$}\label{761610} 

Summarizing all the $k$ on both sides of Eq. \eqref{761533}, we have 
\begin{align}
\sum\limits_{k=0}^{K}\sum\limits_{w=1}^W \eta_w y_{k,w} 
=\sum\limits_{k=0}^{K}\sum\limits_{m=0}^{M-1} \pi_{k-m}\sum\limits_{i=m+1}^{M}\theta_i =\sum\limits_{m=0}^{M-1} \sum\limits_{i=m+1}^{M}\theta_i    =\bar{a},
\end{align}
where the second equality holds because the normalization of the steady-state probabilities, i.e., $\sum\nolimits_{k=0}^{K} \pi_k =1$.

To construct the LP problem \eqref{tradeoff2}, we need to express $\pi_k$ as a linear function of variables $\{y_{k,m}\}$. For ease of exposition, we introduce a $(K\!+\!1) \!\times\! \big[W(K\!+\!1)\big]$ constant matrix $\bm{G}$ to describe the relationship between $\{\pi_k\}$ and  $\{y_{k,m}\}$ based on Eq. \eqref{new-varable}. The $(k+1)$-th row vector of matrix $\bm{G}$, denoted by $g_{k+1}$, is given as
\begin{align}
\left\{
\begin{array}{cll}
\bm{g}_{k+1} &= \frac{1}{r_0} \bm{l}_{1}, & k=0,   \\
\bm{g}_{k+1} &= \frac{1}{r_0} ( \bm{l}_{k+1} - \!\! \sum\limits_{i=1}^{M-1} r_i \bm{g}_{k-i}), & k \in \mathbb{K}. \label{GGneration}
\end{array}
\right.
\end{align}
In Eq. \eqref{GGneration}, $\bm{l}_{k+1}$ is a $W(K\!+\!1)$-dimensional vector whose $(Wk+w)$th element is $\eta_w$ while the other elements are zero. In this way, the steady-state probabilities $\{\pi_k\}$ can be linearly expressed by the set of variables $\{y_{k,w}\}$, namely,
\begin{align}\label{11201516}
\pi_k = \sum_{i=0}^{K}\sum_{j=1}^{W}  G_{(k+1,iW+j)} \cdot y_{i,j},
\end{align}
where $G_{(i,j)}$ is the $(i \times j)$-th element of matrix $\bm{G}$.

\subsubsection{Constraints (\ref{tradeoff1}.c) and (\ref{tradeoff1}.e) can be left out in the new equivalent LP problem}\label{11201529}

The constraint (\ref{tradeoff1}.c) is incorporated in the derivation of Eq. \eqref{11201516} which is the reason that it can be left out. As for constraint (\ref{tradeoff1}.e), since the elements in matrix $\bm{G}$ and variable $y_{i,j}$ are non-negative, as given by Eq. \eqref{11201516}, $\pi_k$ is clearly non-negative. On the other hand, since $r_i$ (c.f.  Eq. \ref{761533}) is non-negative and $\sum\limits_{w=1}^W \eta_w y_{k,w}$ is equal to or less than one, $\pi_k$ is no more than one. Thus, constraint (\ref{tradeoff1}.e) is also implied in  constraints (\ref{tradeoff2}.c). 



In summary, we have shown that how to construct the LP problem \eqref{tradeoff2} step by step equivalently from problem \eqref{tradeoff1}, which completes the proof of \emph{Theorem} \ref{theorem3}.

\section{The proof of \emph{Theorem} \ref{8171101}}\label{appA}

Suppose $\{y_{k,w}\}$ is a set of variables that  minimize the average delay $D$ under constraints \big[\ref{tradeoff2}.(a-c)\big]. The corresponding transmission power $P$ is equal to  $\sum\nolimits_{k=0}^{K}\sum\nolimits_{w=1}^W \eta_w P_w y_{k,w}$. We show that if there exists another set of variables $\{\hat{y}_{k,w}\}$ which cost a higher power $\hat{P} > P$ to transmit, a smaller queueing delay $\hat{D} < D$ will be induced.

Let $k_{th}$ be a positive integer. We construct variables $\{\hat{y}_{k,w}\}$ as follows:
\begin{align}
\left\{
\begin{array}{ll}
\hat{y}_{k,w} = y_{k,w} + \Delta y_{k,w}, & k \leqslant k_{th};\\
\hat{y}_{k,w} = y_{k,w} - \Delta y_{k,w}, & k > k_{th}, 
\end{array}
\right.
\end{align} 
where $\{\Delta y_{k,w}\}$ are set as non-negative quantities to make sure that variables $\{\hat{y}_{k,w}\}$ meet constraints \big[\ref{tradeoff2}.(b-c)\big] and the constraint $\hat{P}-P>0$ is satisfied. Let $\Delta P = \hat{P}-P$, we then have
\begin{align}\label{751518}
\Delta P = \sum\limits_{w=1}^W \eta_w P_w \big(\sum\limits_{k=0}^{k_{th}}\Delta y_{k,w} - \sum\limits_{k=k_{th}+1}^{K}\Delta y_{k,w}\big) > 0,
\end{align}
which means that $\Delta y_{k,w}$ strictly stays positive for some $k \leqslant k_{th}$. 

In this way, the average delay gap satisfies the following inequality 
\begin{align}\label{751546}
&\hat{D}-D \\&= \frac{1}{\bar{a}^2}\sum\limits_{k=0}^{K}\sum\limits_{w=1}^{W}k\eta_w (\hat{y}_{k,w}-y_{k,w})\\
&=\frac{1}{\bar{a}^2}\sum\limits_{w=1}^{W}\eta_w (\sum\limits_{k=0}^{k_{th}}k \Delta y_{k,w} - \sum\limits_{k=k_{th+1}}^{K} k \Delta y_{k,w})\\
&<\frac{1}{\bar{a}^2}\sum\limits_{w=1}^{W}\eta_w[k_{th}\sum\limits_{k=0}^{k_{th}} \Delta y_{k,w}-(k_{th}+1)\sum\limits_{k=k_{th}+1}^{K} \Delta y_{k,w}].\nonumber
\end{align} 

Since both $\{\hat{y}_{k,w}\}$ and $\{{y}_{k,w}\}$ meet constraint $(\ref{tradeoff2}.b)$, we have
\begin{align}\label{751545}
\sum\limits_{w=0}^{W} \eta_w \sum\limits_{k=0}^{k_{th}}\Delta y_{k,w}=\sum\limits_{w=0}^{W} \eta_w \sum\limits_{k=k_{th}+1}^{K}\Delta y_{k,w}.
\end{align}
Combining Eq. \eqref{751546} and Eq. \eqref{751545}, we know $\hat{D}<D$.

We have proven that if $P_{aver}^{'}<P_{aver}^{''}$, then $d(P_{aver}^{''})<d(P_{aver}^{'})$. Thus, the delay-power tradeoff function $d(\cdot)$ is a monotonically decreasing function of the average power.

\section{The proof of \emph{Lemma} \ref{6291113}}\label{proof6291113}

We adopt the proof by contradiction, namely, if there exists a set of optimal variables $\{y_{k,w}\}$ that do not meet Eq. \eqref{629947}, we can find another set of variables ${\{\hat{y}_{k,w}\}}$ that can use less power to obtain the same queueing delay, which contradicts \emph{Theorem} \ref{8171101}. 

Suppose that, there exists a set of optimal variables $\{y_{k,w}\}$, in which there are $0 < w_1 < w_2 \leqslant W$ that make $y_{k,w_1} > y_{k,w_2}$. With this solution, the minimum delay $D$ can be achieved at the power cost of $\Gamma_k$. We construct another set of ${\{\hat{y}_{k,w}\}}$ as
\begin{align}
\hat{y}_{k,w} = \left\{
\begin{array}{ll}
y_{k,w}, & w > w_2\\
y_{k,w_1}, & w = w_2\\
y_{k,w}, & w_1 < w < w_2\\
y_{k,w} - \Delta y_{k,w}, & w \leqslant w_1,    
\end{array}
\right.
\end{align}
where the quantities \{$\Delta y_{k,w}, w \leqslant w_2$\} are all non-negative reals that meet the constraints that $\sum\limits_{w=1}^{w_1} \eta_w \Delta y_{k,w} = \eta_{w_2}(y_{k,w_1}-y_{k,w_2})$ and $0 \leqslant \hat{y}_{k,w} \leqslant \hat{y}_{k,w+1}$. Thus, all the variables  $\hat{y}_{k,w}$ satisfy Eq. \eqref{629947} and  $\sum\nolimits_{w=1}^{W}\eta_w\hat{y}_{k,w} = \sum\nolimits_{w=1}^{W}\eta_w{y}_{k,w}$. Also, we have  $\mathop{\text{max}}\limits_{w}\{\hat{y}_{k,w}\} = \mathop{\text{max}}\limits_{w} \{{y}_{k,w}\}$. Thus, by introducing the new set of variables, the objective function in problem \eqref{tradeoff2} does not change. The corresponding power consumption can be calculated as
\begin{align}
\hat{\Gamma}_k = \sum\nolimits_{w=1}^W{\eta_w}P_w\hat{y}_{k,w} 
& = \Gamma_k+\eta_{w_2}P_{w_2}(y_{k,w_1}-y_{k,w_2}) - \sum\nolimits_{w=1}^{w_1}\eta_wP_w\Delta y_{k,w} \nonumber\\
& < \Gamma_k + P_{w_2}\Big[\eta_{w_2}(y_{k,w_1}-y_{k,w_2}) - \sum\nolimits_{w=1}^{w_1}\eta_w\Delta y_{k,w}\Big]                                                                                                        , \nonumber
\end{align} 
where the last inequality holds due to that less power is consumed for a better channel, namely, $P_{1} \!> \!P_{2}\! >\! \cdots \!>\! P_W$. Thus, we obtain that $\hat{\Gamma_k}<{\Gamma_k}$ since $\sum\nolimits_{w=1}^{w_1} \eta_w \Delta y_{k,w} = \eta_{w_2}(y_{k,w_1}-y_{k,w_2})$. This  means the new constructed variables lead to the same queueing delay while consuming less power. This contradicts the assumption that variables $\{y_{k,w}\}$ are the optimal solution.

\section{The proof of \emph{Lemma} \ref{6291444}}\label{proof6291444}

Let us denote $L = \sum\limits_{k = 0}^{K}k \sum\limits_{m=0}^{M} \theta_{m} \pi_{k\!+\!1\!-\!m}$. With some transformation, we have  
\begin{align}
L& = \theta_0 \sum\limits_{k = 0}^{K}k \pi_{k+1} + \theta_1 \sum\limits_{k = 0}^{K}k \pi_{k} + \cdots +\theta_{M} \sum\limits_{k = 0}^{K}k \pi_{k+1-M} \nonumber\\
& = \theta_0[Q-(1-\pi_0)] +\theta_1Q +\theta_2(Q+1) + \cdots + \theta_M(Q+M-1) \nonumber \\
& = Q(\theta_0+\cdots\theta_M) - \theta_0(1-\pi_0) + \sum\limits_{m=1}^{M-1}m \theta_{m+1}  \nonumber \\
& = Q + \theta_0\pi_0 -\theta_0+\sum\limits_{m=1}^{M-1}m \theta_{m+1}  \nonumber\\
&=  Q + \theta_0\pi_0 + \varsigma,
\end{align}
where $\varsigma = \sum\nolimits_{m=1}^{M-1}m \theta_{m+1}$. Thus, the average queue length $Q$ can be expressed as $Q = L - \theta_0\pi_0 -  \varsigma$. From the Little's Law, we have 
\begin{align}
D = \frac{1}{\bar{a}}(L - \theta_0\pi_0 -  \varsigma).
\end{align}
In Eq. \eqref{11201019}, $y_{k,w}$ satisfies the inequalities  $0 \leqslant y_{k,w} \leqslant \sum\limits_{m=0}^{M} \theta_{m} \pi_{k\!+\!1\!-\!m}$. Thus, we know $\mathop{\text{max}}\limits_{w} \{y_{k,w}\} \leqslant \sum\limits_{m=0}^{M} \theta_{m} \pi_{k\!+\!1\!-\!m}$. Recalling that $L = \sum\limits_{k = 0}^{K}k \sum\limits_{m=0}^{M} \theta_{m} \pi_{k\!+\!1\!-\!m}$, we have $L \geqslant \sum\limits_{k=0}^{K}  k \cdot \mathop{\text{max}}\limits_{w} \{y_{k,w}\}$. Thus,
\begin{align}
D \geqslant  \frac{1}{\bar{a}}(\sum\limits_{k=0}^{K}  k \cdot \mathop{\text{max}}\limits_{w} \{y_{k,w}\} - \theta_0\pi_0 -  \varsigma),
\end{align}
the "=" holds if and only if $\mathop{\text{max}}\limits_{w} \{y_{k,w}\} = \sum\nolimits_{m=0}^{k+1} \theta_{m} \pi_{k\!+\!1\!-\!m}$.

\section{The proof of Theorem \ref{6291439}} \label{proof6291439}

In \emph{Theorem} \ref{6291444}, we have shown that problem \eqref{tradeoff2} is equivalent to problem \eqref{tradeoff3}. To minimize the average queueing delay in \eqref{tradeoff3}, the weighted sum of $\mathop{\text{max}}\limits_w\{y_{k,w}\}$ should be minimized while $\pi_0$ should be maximized. To minimize $\sum\limits_{k=1}^K k \mathop{\text{max}}\limits_w\{y_{k,w}\}$, the term $\mathop{\text{max}}\limits_w\{y_{k,w}\}$ should be assigned its maximum for smaller $k$ and its minimum for larger $k$. Subject to constraint (\ref{tradeoff2}.c), we have
\begin{align}
\mathop{\text{max}}\limits_w\{y_{k,w}^*\}=\left\{
\begin{array}{ll}
\sum\limits_{m=0}^{M}\theta_m\pi_{k+1-m}^*, & k < k^*;\\
0, & k > k^*, 
\end{array}
\right.
\end{align} 
where $k^*$ is a threshold imposed on the queue length. Once the queue length exceeds $k^*$, the maximum of variable $\{y_{k,w}^*\}$ is zero, and hence all of the variables $\{y_{k,w}^*\}$ are equal to zero. Thus, we only consider the situation when $k<k^*$.

Based on the result in \emph{Lemma} \ref{6291113}, we know that for any $k$, there exists $y_{k,1} \leqslant y_{k,2} \leqslant  \cdots \leqslant y_{k,W}$. Thus, $y_{k,W} = \mathop{\text{max}}\limits_w\{y_{k,w}^*\}$, the threshold value $T_k$ introduced in \emph{Theorem} \ref{6291439} meets the constraint that $T_k<W$. Suppose that there exists a set of optimal variables $\{y_{k,w}^*\}$ which violate the assignment described in Eq. \eqref{6291526}, we can always find another set of variables $\{\hat{y}_{k,w}\}$ which consume less power while achieving to achieve the same delay.

Suppose that $\{y_{k,w}^*\}$ are the optimal variables and satisfy Eq. \eqref{6291526}. Accordingly, the steady-state probabilities $\pi_k^*$ are  given by Eq. \eqref{pi-y}, and the minimum delay $D^*$ ia achieved at the cost of power $\Gamma_k$. For a given queue state $k$, we reassign $\hat{\pi}_k=\pi_k^*$ to a new set of variables $\{\hat{y}_{k,w}\}$ as 
\begin{align}
\left\{
\begin{array}{ll}
\hat{y}_{k,w} = 0, \quad\quad  1\leqslant w < T_k-1,\\ 
0 \leqslant \hat{y}_{k,T_k-1} \leqslant \hat{y}_{k,T_k} \leqslant \sum\limits_{m=0}^{M}\theta_m\pi_{k+1-m}^*,\\
\hat{y}_{k,w}=\sum\limits_{m=0}^{M}\theta_m\pi_{k+1-m}^*, \quad T_k<w\leqslant W.
\end{array}
\right.
\end{align} Notice that this set of variables violate the constraints in Eq. \eqref{6291526} since it has more than one term that lies between the maximum and the minimum of $\{\hat{y}_{k,w}\}$. Since the steady-state probabilities do not change, the minimum queueing delay remains the same. By comparing $\{\hat{y}_{k,w}\}$ with $\{y_{k,w}^*\}$, we get $y_{k,T_k-1}^*<\hat{y}_{k,T_k-1}\leqslant \hat{y}_{k,T_k}<y_{k,T_k}^*$. Since $\pi_k^*=\hat{\pi}_k$, from Eq. \eqref{761533}, we have  $\eta_{T_k-1}(y_{k,T_k-1}^*-\hat{y}_{k,T_k-1}) + \eta_{T_k}(y_{k,T_k}^*-\hat{y}_{k,T_k}) = 0$.The power consumption can be calculated as
\begin{align}\label{6291642}
\hat{\Gamma}_k 
&= \sum\limits_{w=1}^W{\eta_w}P_w\hat{y}_{k,w} \nonumber \\
& = \Gamma_k+\eta_{T_k-1}P_{T_k-1}(\hat{y}_{k,T_k-1}-y_{k,T_k-1}^*)  + \eta_{T_k}P_{T_k}(\hat{y}_{k,T_k}-y_{k,T_k}^*) \\
&=\Gamma_k + P_{T_k-1}\eta_{T_k}(y_{k,T_k}^*-\hat{y}_{k,T_k}) + \eta_{T_k}P_{T_k}(\hat{y}_{k,T_k}-y_{k,T_k}^*) \nonumber \\
&=\Gamma_k + \eta_{T_k}(y_{k,T_k}^*-\hat{y}_{k,T_k})(P_{T_k-1}-P_{T_k}) >\Gamma_k. \nonumber
\end{align}  
The last inequality holds since less power is consumed in a better channel, i.e.,  $P_{T_k-1}>P_{T_k}$ and $y_{k,T_k}^*>\hat{y}_{k,T_k}$. This means that $\{y_{k,w}\}$ violating Eq. \eqref{6291526} will cause a higher power consumption. The same conclusion can be obtained when more than two variables violate Eq. \eqref{6291526}. In this way, we prove \emph{Theorem} \ref{6291439} by contradiction.

\section{The proof of \emph{Lemma} \ref{for_w}} \label{prooffor_w}

Given channel state $w$, we define the delay minimization problem subject to the power constraint from the perspective of buffer occupation and power consumption costs discussed in Section \ref{sec4} as follows:
\begin{align}
&\mathop{\text{min}}\limits_{\{s[n]\}} \ \frac{1}{\bar{\alpha}}  \mathbb{E}[(t[n]-s[n]+a[n+1])^+] \label{tradeoff4}\\
&\text{s.t.}  \quad \mathbb{E}(P_w s[n]) \leqslant P_{aver}^{'}, \label{8171557}
\end{align}
where $t[n] = q[n-1] + a[n]$ is the queue state after a new datat arrival. The objective is derived  based on the Little's Law and the average symbol $\mathbb{E}$ means the expectation taken over all the time slots. In the above optimization problem, the scheduling policy is described by the transmission variable $\{s[n]\}$. We mainly consider the optimal scheduling policy in the case when the equality \eqref{8171557} holds. 
If the power constraint is sufficiently large, i.e., the inequality in \eqref{8171557} holds, we only need to apply the scheduling policy to transmit packets as long as the queue is not empty, regardless of the channel state. Thus, using the method of Lagrangian multipliers, we only need to minimize the following Lagrangian function
\begin{align}
\mathop{\text{min}}\limits_{\{s[n],\beta\}} \   L(s[n],\beta') = & \frac{1}{\bar{\alpha}}\mathbb{E}\big[(t[n]-s[n]+a[n+1])^+\big] + \beta'\big[\mathbb{E}(P_w s[n]) - P_{aver}^{'}\big], 
\end{align}
where  $\beta'$ is a Lagrangian multiplier. Equivalently, for each multiplier $\beta'$, we should solve the following problem
\begin{align}\label{8291421}
\mathop{\text{min}}\limits_{\{s[n]\}} \   \mathbb{E}[(t[n]-s[n]+a[n+1])^+] + \beta\mathbb{E}(s[n]),
\end{align}
where $\beta$ is defined as $\bar{\alpha}\beta^{'}P_w$. In the sequel, we show that the optimal solution to Eq. \eqref{8291421} has a threshold structure. We define the expected total cost function in time slot $n$ as
\begin{align}\label{8291509}
C_n(t[n],s[n]) = \mathbb{E}[(t[n]-s[n]+a[n+1])^+] + \beta s[n],
\end{align}
where the first term represents the queue length cost and the second term represents the power cost, respectively. Define $V_n(t[n])$ as the total cost spent from slot $n$ to slot $N$ if we follow the optimal policy from slot $n$ thereafter, namely,
\begin{align}\label{10021610}
V_n(t[n]) =  \mathop{\text{inf}}\limits_{s[n^{'}],n^{'}\geqslant n} \mathbb{E}\sum\limits_{n'=n}^{N}
\gamma^{n'}C_{n'}\big(t[n'],X^{\Omega}(t[n'])\big).
\end{align}
The factor $\gamma$ in Eq.\eqref{10021610} is the discount factor. Hence, the quantity $V_n$ adds up all the costs spent across the slots $n$, $n+1$, $\cdots$, $N$ and ignores the costs spent previously before slot $n$.

Let $z = t[n]-s[n]$. Define $G_n$ as a function of $z$
\begin{align}
G_n(z) = -\beta z + \mathbb{E}[(z+a[n+1])^+] + \gamma \mathbb{E}[V_{n+1}(z+a[n+1])] .
\end{align}
The cost $V_n$ can be rewritten as 
\begin{align}
V_n(t[n]) = \beta t[n] + \mathop{\text{min}}\limits_z G_n(z).
\end{align}
Let $I_w = \mathop{\text{argmin}}\limits_{z}G_n(z)$. Since\footnote{See subsection \ref{convex1}.} $G_n(z)$ is a convex function of $z$ among all the feasible $z$, we should choose a solution $z=I_w$ rather than any other $s[n]$. Otherwise, the transmission action $s[n]$ should always be chosen to make $z=t[n]-s[n]$ approach $I_w$. To this end, when $t[n]$ is greater than $I_w$, the transmission action $s[n]=1$ should be made and otherwise $s[n]=0$ is selected. Combining these two cases, we show that $s[n]$ is obtained as given by Eq. \eqref{12241719}.

\subsection{Function $G_n(z)$ is convex of $z$ for all $n$}\label{convex1}
We use the mathematical induction to prove the convexity of $G_n(z)$. Firstly, considering the fact that term  $(z+a[N+1])^+$ is convex in $z$ for any value of $a[n+1]$,
\begin{align}
G_N(z) = -\beta z + \mathbb{E}[(z+a[N+1])^+]
\end{align}
is convex in $z$. Secondly, we assume that $G_{n+1}(z)$ is convex. Then, we know 
\begin{align}
V_{n+1}(t[n+1]) = \beta t[n+1] + \mathop{\text{min}}\limits_z G_{(n+1)}(z)
\end{align}
is convex. Finally, We can derive $G_n(z)$ as
\begin{align}
G_n(z) = &-\beta z + \mathbb{E}[(z+a[n+1])^+] + \gamma \mathbb{E}[V_{n+1}(z+a[n+1])]
\end{align}
is convex in $z$.

\section{The proof of (2) in \emph{Theorem} \ref{6292306}} \label{nonzero_probability}

\begin{figure*}[t]
	\centering
	\captionsetup[subfigure]{labelformat=simple,captionskip=6bp,nearskip=6bp,farskip=0bp,topadjust=0bp}
	\renewcommand{\thesubfigure}{(\alph{subfigure})}
	\subfigure[Threshold points: marked in grey or red shadow]{
		\includegraphics[width=0.45\columnwidth]{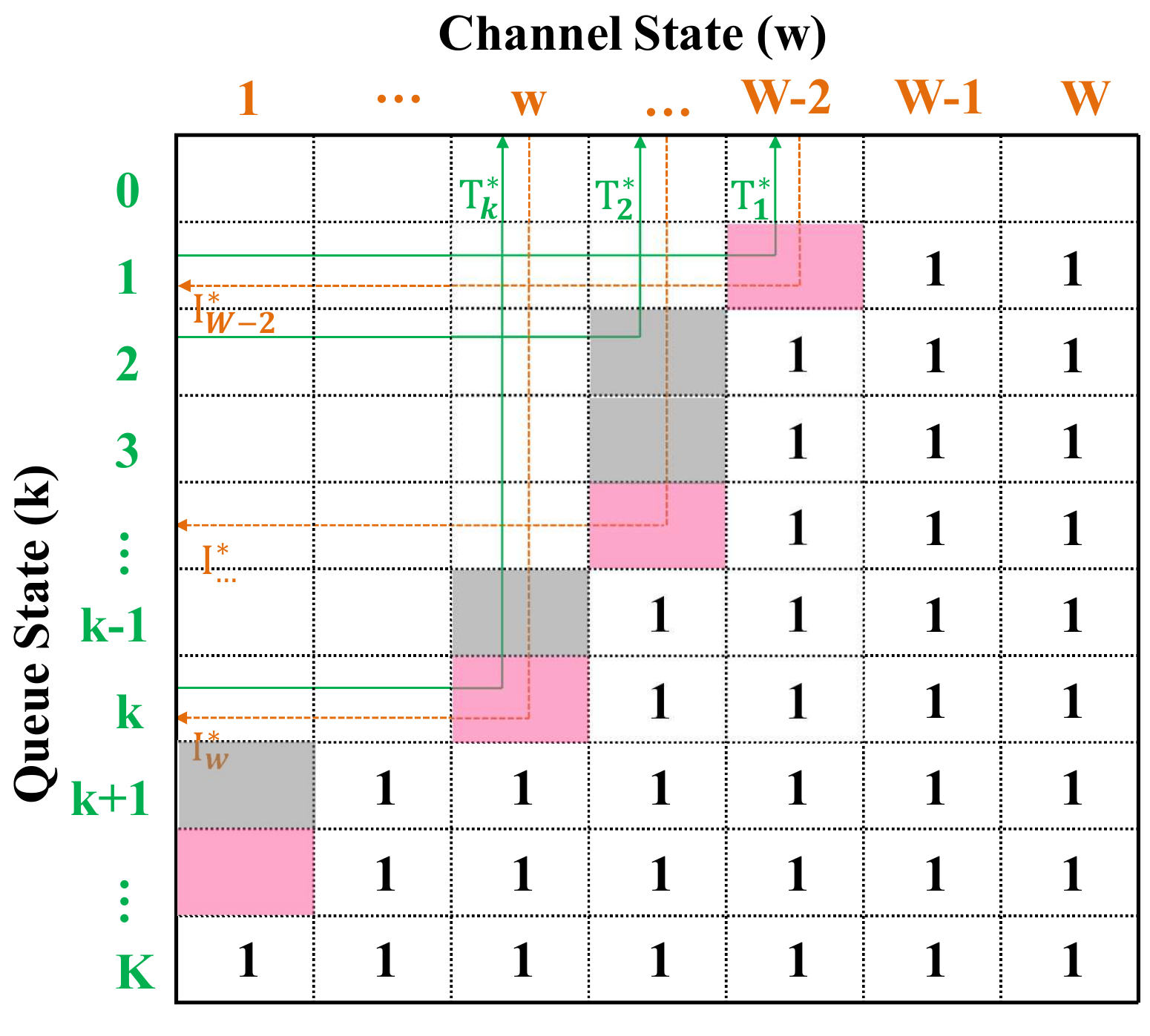} \label{non1}}
	\hspace*{-0.3cm}
	\subfigure[At most one nonzero probability at the threshold points]{
		\includegraphics[width=0.45\columnwidth]{figure/dual_threshold_2} \label{non2}}
	\caption{An illustration of conclusion (2) in \emph{Theorem} \ref{6292306}}\label{non}
	\vspace{-0.0cm}
\end{figure*}

As shown in the \figurename\ref{non1}, the threshold points are highlighted with the shadow rectangles (in red and gray). We first notice that some of the values on the threshold points are zero due to the threshold structure imposed on the queue length (the vertical direction), like the gray areas. Then, we show that there is at most one nonzero value for these left threshold points in what follows.
	
We use contradiction proof. Suppose that $\{y_{k,w}^*\}$ is the optimal solution for a given power constraint $P^*$, and there are two threshold points $(k_1, w_1)$ and $(k_2, w_2 )$, whose scheduling probabilities  are nonzero. If $\frac{k_1}{k_2} \geqslant \frac{P_{w_1}}{P_{w_2}}$, we can reconstruct a new set of $\{y_{k,w}^{'}\}$ by reducing $y_{k_1,w_1}^*$ and increasing $y_{k_2,w_2}^*$ based on equation $\eta_{w_1}P_{w_1}\Delta y^{*}_{k_1,w_1} = \eta_{w_2}P_{w_2}\Delta y^{*}_{k_2,w_2}$, until $y_{k_1,w_1}^*$ is zero or $y_{k_2,w_2}^*$ becomes its maximum as given in Eq. (23)
. Let $D^{'}$ and $P^{'}$ denote the average delay and power induced under parameters $\{y_{k,w}^{'}\}$. Based on the expression of the average power in Eq. (19), the difference between $P^{'}$ and $P^*$ is 
\begin{align}
\Delta P = P^{'} -  P^* =  \eta_{w_2}P_{w_2}\Delta y^{*}_{k_2,w_2} - \eta_{w_1}P_{w_1}\Delta y^{*}_{k_1,w_1}  = 0,
\end{align}
and the difference between $D^{'}$ and $D^*$
\begin{align}
\Delta D &= D^{'} -  D^* \nonumber\\ &= \frac{1}{\bar{a}^2} \left( k_2\eta_{w_2}\Delta y^{*}_{k_2,w_2} - k_1 \eta_{w_1}\Delta y^{*}_{k_1,w_1} \right) \nonumber\\
&= \frac{\eta_{w_1} \Delta y^{*}_{k_1,w_1} }{\bar{a}^2 P_{w_2}} \left( k_2P_{w_1} - k_1 P_{w_2} \right) \leqslant 0.
\end{align}
It means that, with the new constructed parameters, we can obtain a lower average delay which contradicts the optimality of the $\{y_{k,w}^*\}$. Thus, there is at most one threshold point at which $y_{k,w}^*$ takes value between zero and its maximum. For the case  $\frac{k_1}{k_2} < \frac{P_{w_1}}{P_{w_2}}$, same conclusion can be obtained if we reduce $y_{k_2,w_2}^*$ and increase $y_{k_1,w_1}^*$. If there are more than two nonzero threshold points, we can eliminate the nonzero values one by one and degrade to the above case.

Based on the relationship between$\{y_{k,w}^*\}$ and $\{f_{k,w}^*\}$ in Eq. (18), we can obtain the conclusion.

\section{A straightforward way to explain the dual-threshold-based policy } \label{intuition}

\begin{figure*}[b]
	\centering
	\captionsetup[subfigure]{labelformat=simple,captionskip=6bp,nearskip=6bp,farskip=0bp,topadjust=0bp}
	\renewcommand{\thesubfigure}{(\alph{subfigure})}
	\subfigure[Power allocation without considering the channel state]{
		\includegraphics[width=0.45\columnwidth]{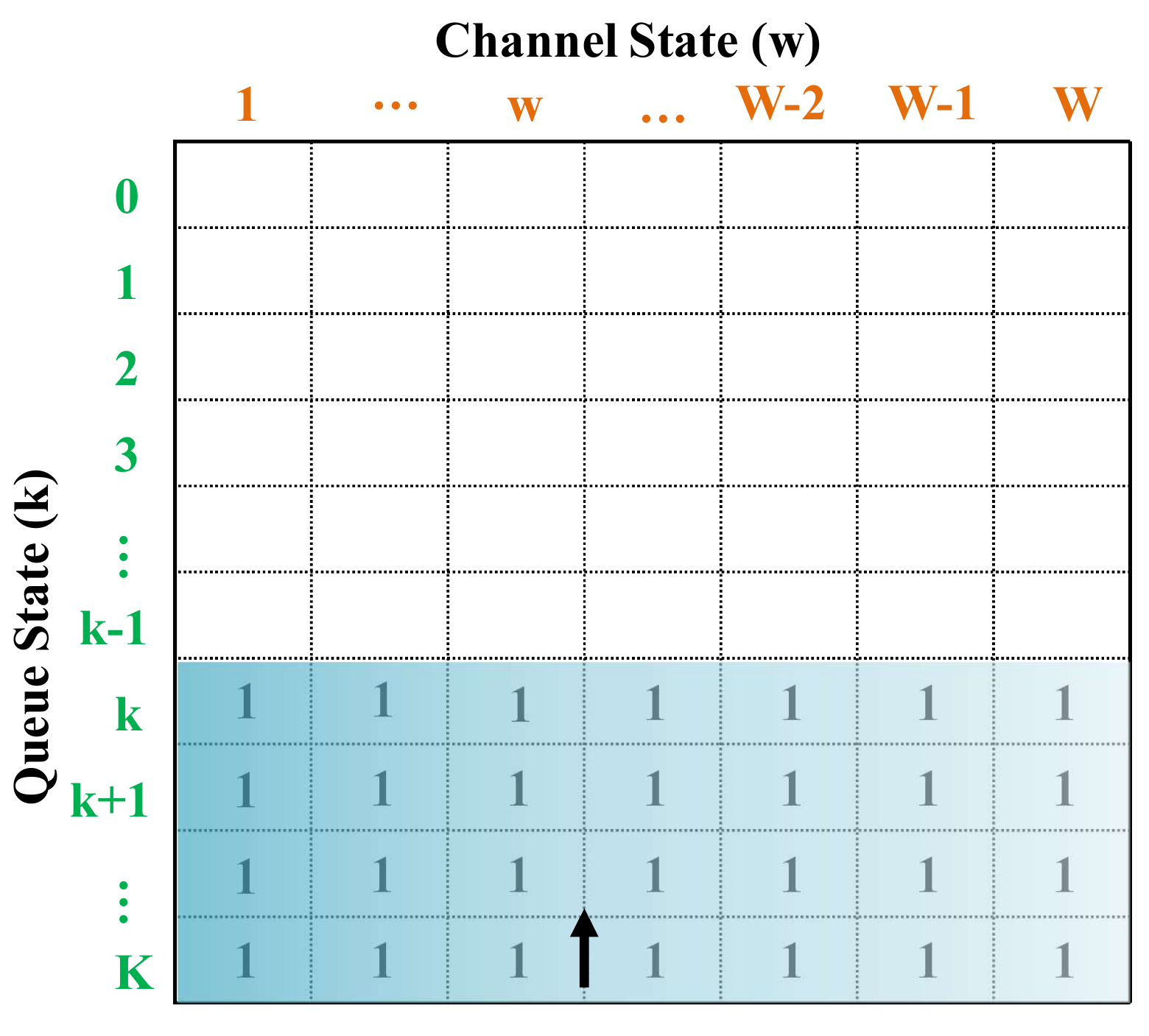} \label{left}}
	\hspace*{-0.3cm}
	\subfigure[Power allocation with joint queue and channel states ]{
		\includegraphics[width=0.45\columnwidth]{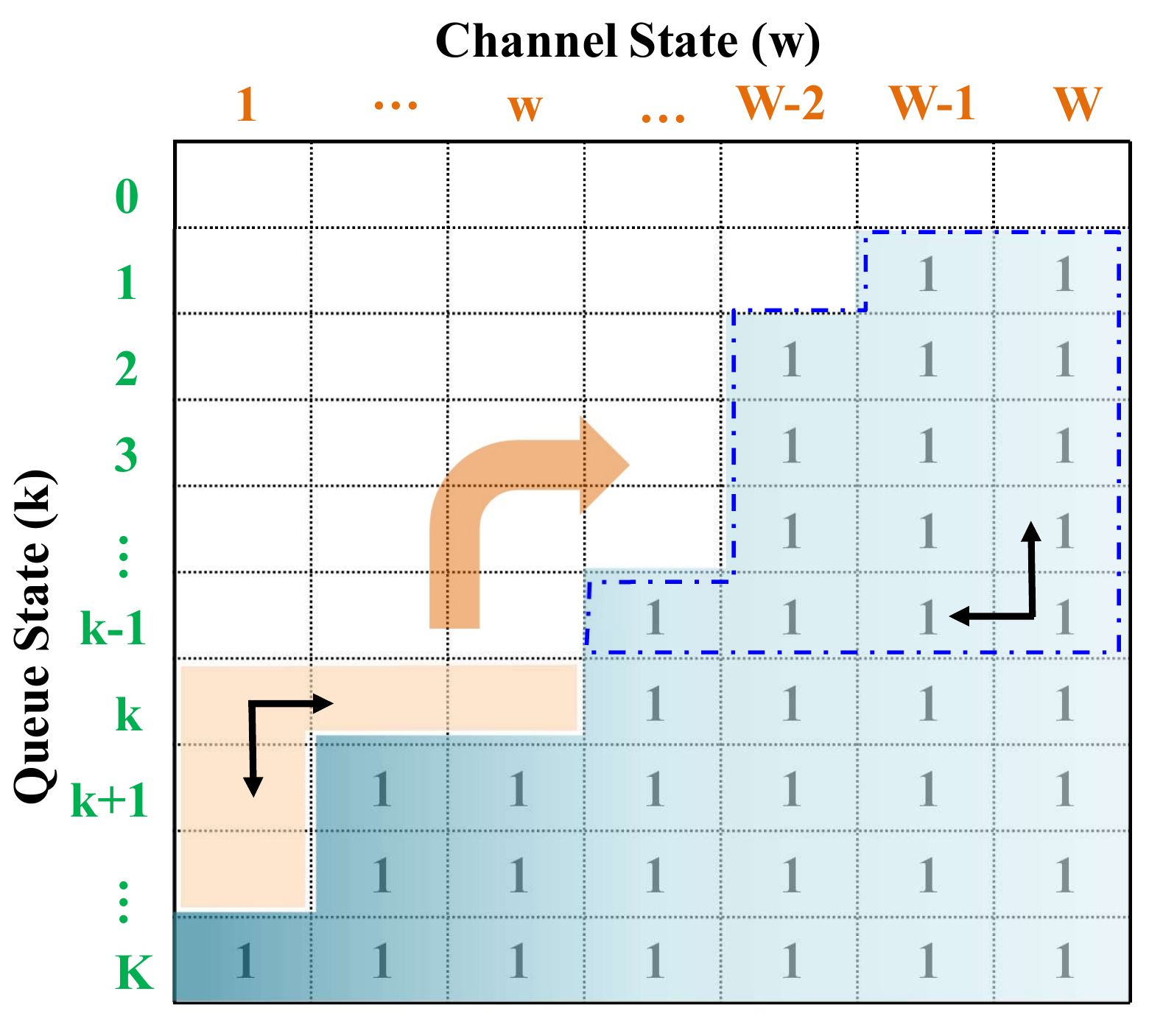} \label{right}}
	\caption{An explanation of the dual-threshold-based policy}\label{614936}
\end{figure*}

For the dual-threshold-based policy, we would like to give a straightforward way to explain why the solution is a threshold policy with the help of the following picture. The average delay has the form of weighted sum with the steady-state probability $\pi_k$ being the weight, i.e., $D = \frac{1}{\bar{a}}\sum\limits_{k=0}^{K}k \pi_k$. In order to minimize the average delay, the limit power resource should be allocated to decrease the $\pi_k$ which has greater $k$. As shown in \figurename\ref{left}, the limit power can only afford to transmit for some greater $k$. However, transmitting one packet over a bad channel consumes much more power than over a better channel. Thus, the power consumed over these bad channels for greater $k$ can be moved to transmit for relative lower $k$ but over better channels. In \figurename\ref{right}, we remove the power from the orange area to the blue circled area. The black arrows show the order of priority of the move process. Thus, the optimal policy shows up with the dual-threshold-based policy.

\end{spacing}
%

%
%
%
%
%
%
%

\begin{spacing}{1.15}
\bibliographystyle{IEEEtran}
\bibliography{IEEEabrv,scheduling}

\begin{thebibliography}{10}
\providecommand{\url}[1]{#1}
\csname url@samestyle\endcsname
\providecommand{\newblock}{\relax}
\providecommand{\bibinfo}[2]{#2}
\providecommand{\BIBentrySTDinterwordspacing}{\spaceskip=0pt\relax}
\providecommand{\BIBentryALTinterwordstretchfactor}{4}
\providecommand{\BIBentryALTinterwordspacing}{\spaceskip=\fontdimen2\font plus
\BIBentryALTinterwordstretchfactor\fontdimen3\font minus
  \fontdimen4\font\relax}
\providecommand{\BIBforeignlanguage}[2]{{%
\expandafter\ifx\csname l@#1\endcsname\relax
\typeout{** WARNING: IEEEtran.bst: No hyphenation pattern has been}%
\typeout{** loaded for the language `#1'. Using the pattern for}%
\typeout{** the default language instead.}%
\else
\language=\csname l@#1\endcsname
\fi
#2}}
\providecommand{\BIBdecl}{\relax}
\BIBdecl

\bibitem{6824752}
J.~G. Andrews, S.~Buzzi, W.~Choi, S.~V. Hanly, A.~Lozano, A.~C.~K. Soong, and
  J.~C. Zhang, ``What will 5\protect{G} be?'' \emph{IEEE Journal on Selected
  Areas in Communications}, vol.~32, no.~6, pp. 1065--1082, June 2014.

\bibitem{6157574}
D.~Feng, C.~Jiang, G.~Lim, L.~J. Cimini, G.~Feng, and G.~Y. Li, ``A survey of
  energy-efficient wireless communications,'' \emph{IEEE Communications Surveys
  Tutorials}, vol.~15, no.~1, pp. 167--178, First 2013.

\bibitem{fettweis2014tactile}
G.~P. Fettweis, ``The tactile internet: Applications and challenges,''
  \emph{IEEE Vehicular Technology Magazine}, vol.~9, no.~1, pp. 64--70, 2014.

\bibitem{Popovski2017Ultra}
P.~Popovski, J.~J. Nielsen, C.~Stefanovic, E.~de~Carvalho, E.~Ström, K.~F.
  Trillingsgaard, A.-S. Bana, D.~M. Kim, R.~Kotaba, and J.~Park,
  ``Ultra-reliable low-latency communication (urllc): Principles and building
  blocks,'' 2017.

\bibitem{1365285}
R.~A. Berry, ``Order optimal energy efficient transmission policies in the
  small delay regime,'' in \emph{International Symposium on Information
  Theory}, June 2004.

\bibitem{6802871}
J.~Kakarla and B.~Majhi, ``A new optimal delay and energy efficient
  coordination algorithm for wsan,'' in \emph{2013 IEEE International
  Conference on Advanced Networks and Telecommunications Systems (ANTS)}, Dec
  2013, pp. 1--6.

\bibitem{720543}
A.~Ephremides and B.~Hajek, ``Information theory and communication networks: an
  unconsummated union,'' \emph{IEEE Transactions on Information Theory},
  vol.~44, no.~6, pp. 2416--2434, Oct 1998.

\bibitem{collins1999transmission}
B.~Collins and R.~L. Cruz, ``Transmission policies for time varying channels
  with average delay constraints,'' in \emph{Proc. Allerton Conf.
  Communication, Control, and Computing, Monticello, IL}, 1999, pp. 709--717.

\bibitem{berry2002communication}
R.~A. Berry and R.~G. Gallager, ``Communication over fading channels with delay
  constraints,'' \emph{IEEE Transactions on Information Theory}, vol.~48,
  no.~5, pp. 1135--1149, 2002.

\bibitem{berry2004cross}
R.~A. Berry and E.~M. Yeh, ``Cross-layer wireless resource allocation,''
  \emph{IEEE Signal Processing Magazine}, vol.~21, no.~5, pp. 59--68, 2004.

\bibitem{6482230}
R.~A. Berry, ``Optimal power-delay tradeoffs in fading channels-small-delay
  asymptotics,'' \emph{IEEE Transactions on Information Theory}, vol.~59,
  no.~6, pp. 3939--3952, June 2013.

\bibitem{6152121}
G.~Saleh, A.~El-Keyi, and M.~Nafie, ``Cross-layer minimum-delay scheduling and
  maximum-throughput resource allocation for multiuser cognitive networks,''
  \emph{IEEE Transactions on Mobile Computing}, vol.~12, no.~4, pp. 761--773,
  April 2013.

\bibitem{6477062}
J.~Choi, ``Energy-delay tradeoff comparison of transmission schemes with
  limited csi feedback,'' \emph{IEEE Transactions on Wireless Communications},
  vol.~12, no.~4, pp. 1762--1773, April 2013.

\bibitem{1425747}
H.~Wang and N.~B. Mandayam, ``Opportunistic file transfer over a fading channel
  under energy and delay constraints,'' \emph{IEEE Transactions on
  Communications}, vol.~53, no.~4, pp. 632--644, April 2005.

\bibitem{uysal2002energy}
E.~Uysal-Biyikoglu, B.~Prabhakar, and A.~El~Gamal, ``Energy-efficient packet
  transmission over a wireless link,'' \emph{IEEE/ACM Transactions on
  Networking}, vol.~10, no.~4, pp. 487--499, 2002.

\bibitem{zafer2009calculus}
M.~A. Zafer and E.~Modiano, ``A calculus approach to energy-efficient data
  transmission with quality-of-service constraints,'' \emph{IEEE/ACM
  Transactions on Networking}, vol.~17, no.~3, pp. 898--911, 2009.

\bibitem{5510780}
J.~Yang and S.~Ulukus, ``Delay-minimal transmission for average power
  constrained multi-access communications,'' \emph{IEEE Transactions on
  Wireless Communications}, vol.~9, no.~9, pp. 2754--2767, September 2010.

\bibitem{chen2007optimal}
W.~Chen, Z.~Cao, and K.~B. Letaief, ``Optimal delay-power tradeoff in wireless
  transmission with fixed modulation,'' in \emph{Proc. IEEE IWCLD}, 2007, pp.
  60--64.

\bibitem{6213038}
C.~Isheden, Z.~Chong, E.~Jorswieck, and G.~Fettweis, ``Framework for link-level
  energy efficiency optimization with informed transmitter,'' \emph{IEEE
  Transactions on Wireless Communications}, vol.~11, no.~8, pp. 2946--2957,
  August 2012.

\bibitem{ata2005dynamic}
B.~Ata, ``Dynamic power control in a wireless static channel subject to a
  quality-of-service constraint,'' \emph{Operation Research}, vol.~53, no.~5,
  pp. 842--851, 2005.

\bibitem{1262621}
D.~Rajan, A.~Sabharwal, and B.~Aazhang, ``Delay-bounded packet scheduling of
  bursty traffic over wireless channels,'' \emph{IEEE Transactions on
  Information Theory}, vol.~50, no.~1, pp. 125--144, Jan 2004.

\bibitem{1315903}
H.~Wang and N.~B. Mandayam, ``A simple packet-transmission scheme for wireless
  data over fading channels,'' \emph{IEEE Transactions on Communications},
  vol.~52, no.~7, pp. 1055--1059, July 2004.

\bibitem{7417380}
M.~Wang and W.~Chen, ``Achieving the optimal delay-power tradeoff in wireless
  transmission with arbitrarily random packet arrival: A cross-layer
  approach,'' in \emph{2015 IEEE Global Communications Conference (GLOBECOM)},
  Dec 2014, pp. 1--6.

\bibitem{19920626}
------, ``Delay-power tradeoff of fixed-rate wireless transmission with
  arbitrarily bursty traffics,'' \emph{IEEE Access}, vol.~5, pp. 1668--1681,
  2017.

\bibitem{7893801}
X.~Chen, W.~Chen, J.~Lee, and N.~B. Shroff, ``Delay-optimal buffer-aware
  scheduling with adaptive transmission,'' \emph{IEEE Transactions on
  Communications}, vol.~65, no.~7, pp. 2917--2930, July 2017.

\bibitem{7848871}
M.~Wang, J.~Liu, and W.~Chen, ``Delay optimal scheduling of arbitrarily bursty
  traffic over multi-state time-varying channels,'' in \emph{2016 IEEE Globecom
  Workshops (GC Wkshps)}, Dec 2016, pp. 1--6.

\bibitem{qiao2009impact}
D.~Qiao, M.~C. Gursoy, and S.~Velipasalar, ``The impact of \protect{QoS}
  constraints on the energy efficiency of fixed-rate wireless transmissions,''
  \emph{IEEE Transactions on Wireless Communications}, vol.~8, no.~12, pp.
  5957--5969, 2009.

\bibitem{kleinrock1975queueing}
L.~Kleinrock, ``Queueing systems, volume 1: Theory,'' \emph{Lecture Notes in
  Computer Science}, 1975.

\end{thebibliography}
\end{spacing}

%




\end{document}